\documentclass[12pt]{iopart}
\usepackage{graphicx}

\newcommand{\be}{\begin{eqnarray}}
\newcommand{\ee}{\end{eqnarray}}
\newcommand{\no}{\nonumber}
\def\Vec#1{\mbox{\boldmath $#1$}}

\begin{document}
\title
[Ensemble equivalence in spin systems with short-range interactions]
{Ensemble equivalence in spin systems with short-range interactions}
\author{Kazutaka Takahashi$^1$, Hidetoshi Nishimori$^1$ 
 and Victor Martin-Mayor$^{2,3}$}
\address{
 $^1$
 Department of Physics, 
 Tokyo Institute of Technology, Tokyo 152-8551, Japan

 $^2$
 Departamento de F\'isica Te\'orica I, Universidad Complutense, 
 28040 Madrid, Spain

 $^3$
 Instituto de Biocomputaci\'on y F\'isica de Sistemas Complejos (BIFI), 
 Zaragoza, Spain
}
\begin{abstract}
 We study the problem of ensemble equivalence in spin systems with short-range
 interactions under the existence of a first-order phase transition.
 The spherical model with nonlinear nearest-neighbour interactions is solved
 exactly both for canonical and microcanonical ensembles.
 The result reveals apparent ensemble inequivalence at the first-order
 transition point in the sense that the microcanonical entropy is non-concave
 as a function of the energy and consequently the specific heat is negative.
 In order to resolve the paradox, we show that an unconventional saddle point
 should be chosen in the microcanonical calculation that 
 represents a phase separation.
 The $XY$ model with non-linear interactions is also studied by microcanonical
 Monte Carlo simulations in two dimensions to see how this model behaves
 in comparison with the spherical model.
\end{abstract}

\maketitle

\section{Introduction}

 In statistical mechanics, we prepare an ensemble of macroscopic systems and
 calculate thermodynamic quantities by taking the average over the ensemble.
 When the system is isolated from the environment, 
 the total energy is kept constant
 and the principle of equal weights leads to the microcanonical ensemble.
 On the other hand, when we consider a heat bath attached to the system to allow
 an energy exchange, we have the canonical ensemble characterized by temperature.
 These ensembles are generally considered equivalent and their thermodynamic
 potentials are related by the Legendre transformation.

 Equivalence of ensembles has been proven rigorously for systems with
 short-range interactions \cite{Ruelle}.
 For systems with long-range interactions, there is no guarantee
 that two ensembles produce the same results in the thermodynamic limit.
 Typical examples include gravitational systems \cite{LBW}-\cite{LBLB2}
 and fully-connected mean-field spin models \cite{BMR}-\cite{BN}.
 For a review, see~\cite{CDR}.
 In the latter models, in particular, 
 the interplay of long-range interactions and first-order phase transitions 
 is now known to lead to ensemble inequivalence,
 typically as negative specific heat in the microcanonical ensemble.
 
 In systems with short-range interactions, by contrast, ensembles are equivalent
 in the thermodynamic limit and there should exist no anomalous effects except
 in finite-size systems \cite{finite}.
 In the present paper, we solve the multi-component spin model with nonlinear
 interactions in two and three dimensions exactly for the spherical model
 and numerically for the $XY$ model.
 These models have been known to have first-order phase transitions in two
 and three dimensions \cite{DSS}-\cite{vES2}.
 We show that ensemble equivalence should be taken with special caution 
 in these systems.

 The organization of this paper is as follows.
 In section~\ref{model}, we define the model. 
 The spherical (many-component) limit is solved exactly  in section~\ref{sph}.
 The results for the canonical and microcanonical ensembles are compared. 
 To study the system with finite component spins
 we use microcanonical Monte Carlo simulations in section~\ref{montecarlo}.
 The last section is devoted to summary and conclusion.

\section{$n$-vector model with nonlinear interactions}
\label{model}

 We study the generalized $n$-vector model ($O(n)$-symmetric model)
\be
 H = -Jn\sum_{\langle ij\rangle}V\left(
 \Vec{S}_i\cdot\Vec{S}_j/n\right)  \label{Hamiltonian}
\ee
 on a $d$-dimensional hypercubic lattice.
 The spin variable $\Vec{S}_i$ at site $i$ is an $n$-component vector with the
 constraint $\Vec{S}_i^2=\sum_{a=1}^n(S_i^a)^2=n$.
 The sum in the Hamiltonian is taken over nearest-neighbour pairs.
 The number of spins is $N=L^d$, where $L$ is the linear size of the system.
 In the standard $n$-vector model with linear interactions, 
 the function $V(x)$ is equal to $x$.
 Here, following~\cite{BGH, CP}, we consider the form 
\be
 V(x)= \frac{1}{2^{p-1}p}\left[(1+x)^p-1\right].
\ee
 The linear interaction is recovered if we choose $p=1$.

 The linear model in the limit $n\to \infty$ is the ordinary
 spherical model and can be solved exactly.
 We shall call the nonlinear model also the spherical model for simplicity.
 The canonical analysis of the linear case is found 
 in standard textbooks~\cite{Mussardo, NO}.
 The microcanonical analysis was performed in~\cite{Behringer} and \cite{Kastner}.
 We generalize their calculations to the nonlinear case.

\section{Spherical limit}
\label{sph}

 In the canonical ensemble, the generalized $n$-vector model (\ref{Hamiltonian})
 can be  solved exactly in the spherical limit $n\to\infty$~\cite{CP}.
 The problem is reduced to solving a saddle-point equation for auxiliary variables.
 We solve the nonlinear model in the microcanonical ensemble and 
 compare the results with those for the canonical ensemble.

\subsection{Saddle point equations}

 First, we briefly review how the problem is solved 
 in the canonical ensemble following~\cite{CP}.
 The partition function is written as 
\be
 Z =  \Tr\left\{
 \exp\left[
 \beta Jn\sum_{\langle ij\rangle}V\left(
 \Vec{S}_i\cdot\Vec{S}_j/n\right)
 \right]
 \prod_{i=1}^N\delta\left(\Vec{S}_i^2-n\right)
 \right\},
\ee
 where $\beta$ is the inverse temperature
 and the trace denotes integrations over the spin variables.
 In order to carry out the integrations, the $\delta$ function is 
 expressed by a Fourier integral over the auxiliary variable $z_i$.
 We also introduce two kinds of variables  $\rho_{ij}$
 ($=\Vec{S}_i\cdot\Vec{S}_j/n$) and $\lambda_{ij}$ 
 (to impose the constraint $\rho_{ij}=\Vec{S}_i\cdot\Vec{S}_j/n$)
 and write 
\be
 Z &=&  
 \int \prod_{i=1}^Ndz_i\prod_{\langle ij\rangle}d\lambda_{ij}
 d\rho_{ij}\, \Tr\exp\left[
 \beta Jn\sum_{\langle ij\rangle}V\left(\rho_{ij}\right)
 -\sum_{i}z_i(\Vec{S}_{i}^2-n)
 \right.\no\\
 & & \qquad\qquad\qquad\qquad\left.
 -\sum_{\langle ij\rangle}\lambda_{ij}
 \left(n\rho_{ij}-\Vec{S}_{i}\cdot\Vec{S}_{j}\right)
 \right] \no\\
 &=&  \int\prod_{i=1}^Ndz_i\prod_{\langle ij\rangle}
 d\lambda_{ij} d\rho_{ij}\, \exp\left[
 \beta Jn\sum_{\langle ij\rangle}V\left(\rho_{ij}\right)
 +n\sum_{i}z_i
 -n\sum_{\langle ij\rangle}\lambda_{ij}\rho_{ij}
 \right.\no\\
 & & \qquad\qquad\qquad\qquad \left.
 +n\ln\Tr\exp\left(
 -\sum_iz_iS_i^2+\sum_{\langle ij\rangle}\lambda_{ij}S_iS_j
 \right)\right].
\ee
 The spin trace is just a Gaussian integral 
 over unconstrained scalars $\{S_i\}$ and 
 can be evaluated using the lattice Green function~\cite{Mussardo, NO}. 
 In the limit $n\to\infty$, 
 auxiliary variables are determined from the saddle-point equations.
 Following the conventional procedure used for  
 the spherical model with linear interactions,
 we neglect the subscript dependence of the variables,
 $\lambda=\lambda_i, z=z_i, \rho=\rho_{ij}~(\forall i, j)$.
 Then, we can write 
\be
 Z &=& \exp\left[
 Nnd\beta JV\left(\rho\right)
 +Nn z-Nnd\lambda\rho
 +\frac{n}{2}\sum_{k}\ln
 G(k,\tilde{z})
 \right.\no\\
 & & \qquad 
 -\frac{Nn}{2}\ln\lambda +\frac{Nn}{2}\ln \pi
 \Biggr],
 \label{Z}
\ee
 where $\tilde{z}=z/\lambda$.
 The lattice Green function in the momentum space is given by 
\be
 G(k,\tilde{z}) = \frac{1}{\tilde{z}-\sum_{\mu=1}^d\cos k_\mu}. \label{Green}
\ee
 If we take the thermodynamic limit, the sum over $k$ is replaced by an integral as
\be
 \frac{1}{N}\sum_{k} \to \int \frac{d^dk}{(2\pi)^d}.
\ee

 From the expression~(\ref{Z}), we determine the state of the system 
 by a set of saddle-point equations,
\be
  \lambda = \beta JV'(\rho), \quad 2\lambda = g(\tilde{z}), \quad
 d\rho = \tilde{z}-\frac{1}{2\lambda}, \label{saddle_point_eq_canonical}
\ee
 where 
\be
 g(\tilde{z})= \int \frac{d^dk}{(2\pi)^d}
 \frac{1}{\tilde{z}-\sum_{\mu=1}^d\cos k_\mu}.
 \label{gz}
\ee
 The auxiliary variables $\lambda$ and $\rho$ are eliminated to obtain 
\be 
 \beta J 
 = \frac{g(\tilde{z})}
 {2V'\left(\displaystyle \frac{\tilde{z}}{d}-\frac{1}{dg(\tilde{z})}\right)}.
 \label{spc}
\ee
 For a given $\beta J$, $\tilde{z}$ is determined from this equation.
 Then, the free energy density $f=F/Nn$ is given by
\be
 -\beta f &=& 
 d\beta JV\left(\frac{\tilde{z}}{d}-\frac{1}{dg(\tilde{z})}\right)
 -\frac{1}{2}\ln g(\tilde{z})
 +\frac{1}{2}\int \frac{d^dk}{(2\pi)^d}\ln G(k,\tilde{z})
 \no\\ 
 & & 
 +\frac{1}{2}(1+\ln 2\pi). \label{f}
\ee
 Thus, by solving the simple saddle-point equation~(\ref{spc}) 
 for $\tilde{z}$,  we can calculate
 the free energy as a function of $\beta$.

 Next, we derive the equations in the microcanonical ensemble.
 If we compare (\ref{f}) with the relation $-\beta F=-\beta E+S$,
 we may guess that the energy and entropy are given as
 \be
 & & -\epsilon 
 =dV\left(\frac{\tilde{z}}{d}-\frac{1}{dg(\tilde{z})}\right),
 \label{spmc} \\
 & & s =
 -\frac{1}{2}\ln g(\tilde{z})
 +\frac{1}{2}\int \frac{d^dk}{(2\pi)^d}\ln G(k,\tilde{z})
 +\frac{1}{2}(1+\ln 2\pi),
 \label{s}
\ee
 where $\epsilon=E/NnJ$ and $s=S/Nn$.
 Here, $\tilde{z}$ is obtained as a function of $\epsilon$ by (\ref{spmc})
 to determine the entropy $s=s(\epsilon)$.
 These expressions (\ref{spmc}) and (\ref{s}) for the energy and entropy
 can indeed be derived directly from the microcanonical number of states
 $\Omega = e^S = \Tr \delta(E-H)$ 
 using the integral representation of the delta function. 
 Following the same procedure as in the canonical case, we find
\be
 \Omega (\epsilon)&=& \int \frac{dt}{2\pi}\,\exp\Biggl[
 iNnJt\left\{\epsilon+dV(\rho)\right\} 
 +Nnz-Nnd\lambda\rho
 \no\\
 & & 
 +n\ln\Tr\exp\Biggl(
 -\sum_{i=1}^{N}zS_i^{2}
 +\sum_{\langle ij\rangle}\lambda S_iS_j
 \Biggr)\Biggr].
 \label{Omega}
\ee
 Then, we impose the saddle-point conditions for the auxiliary 
 variables to obtain the above result (\ref{spmc}) and (\ref{s})
 as described in more detail in \ref{app:smicro}.

 We are ready to study the ensemble dependence of system properties
 by comparing the canonical result~(\ref{spc}), (\ref{f})  
 and the microcanonical (\ref{spmc}), (\ref{s}). 
 In the following, we focus ourselves on the cases of $d=2$ and $d=3$.

\subsection{$d=2$}

\begin{center}
\begin{figure}[htb]
\begin{minipage}[h]{0.5\textwidth}
\begin{center}
\includegraphics[width=0.9\columnwidth]{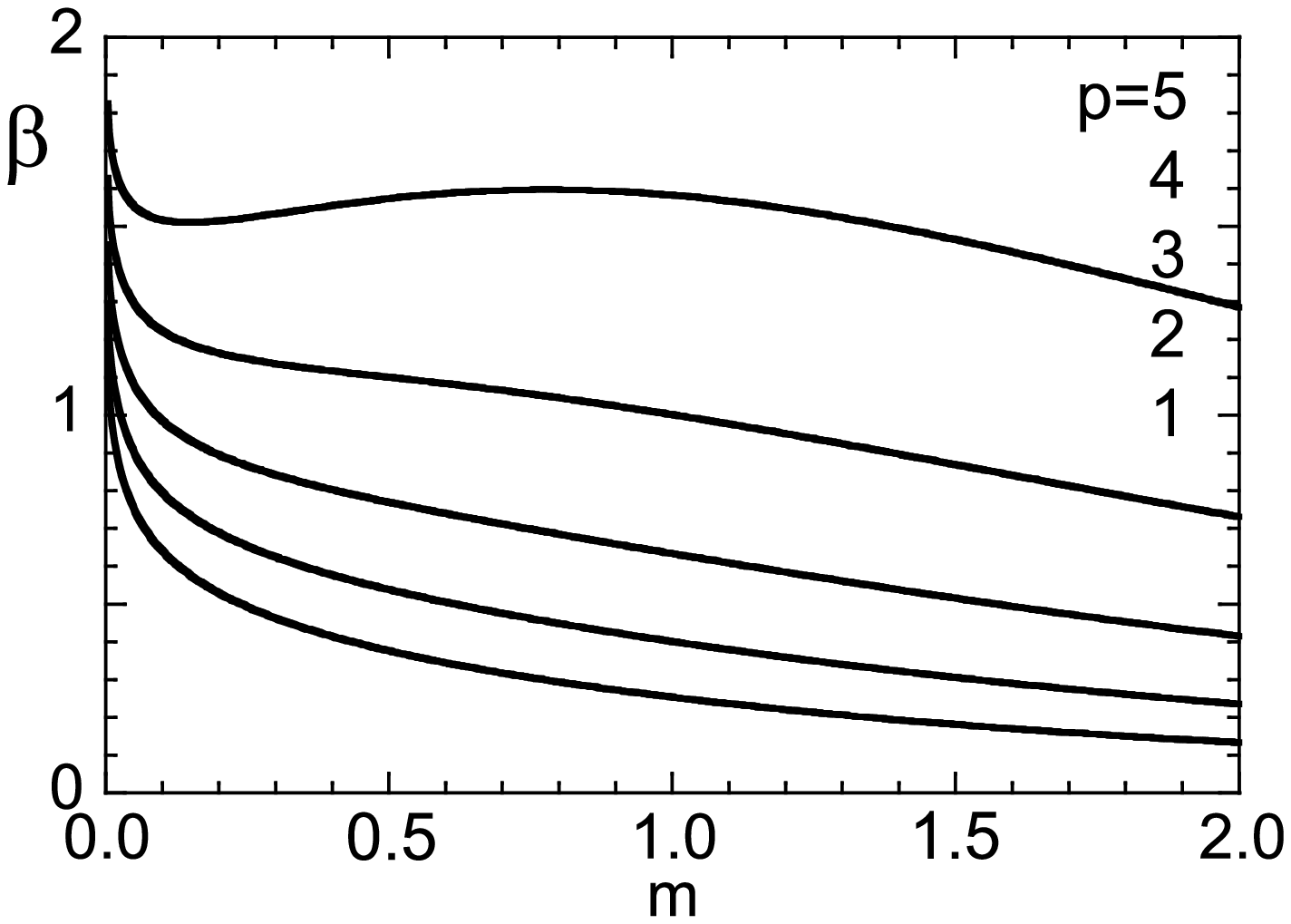}
\caption{Saddle-point equation (\ref{spc})
in the canonical ensemble at $d=2$.
The function diverges logarithmically at the origin.}
\label{sp-c2}
\end{center}
\end{minipage}
\begin{minipage}[h]{0.5\textwidth}
\begin{center}
\includegraphics[width=0.9\columnwidth]{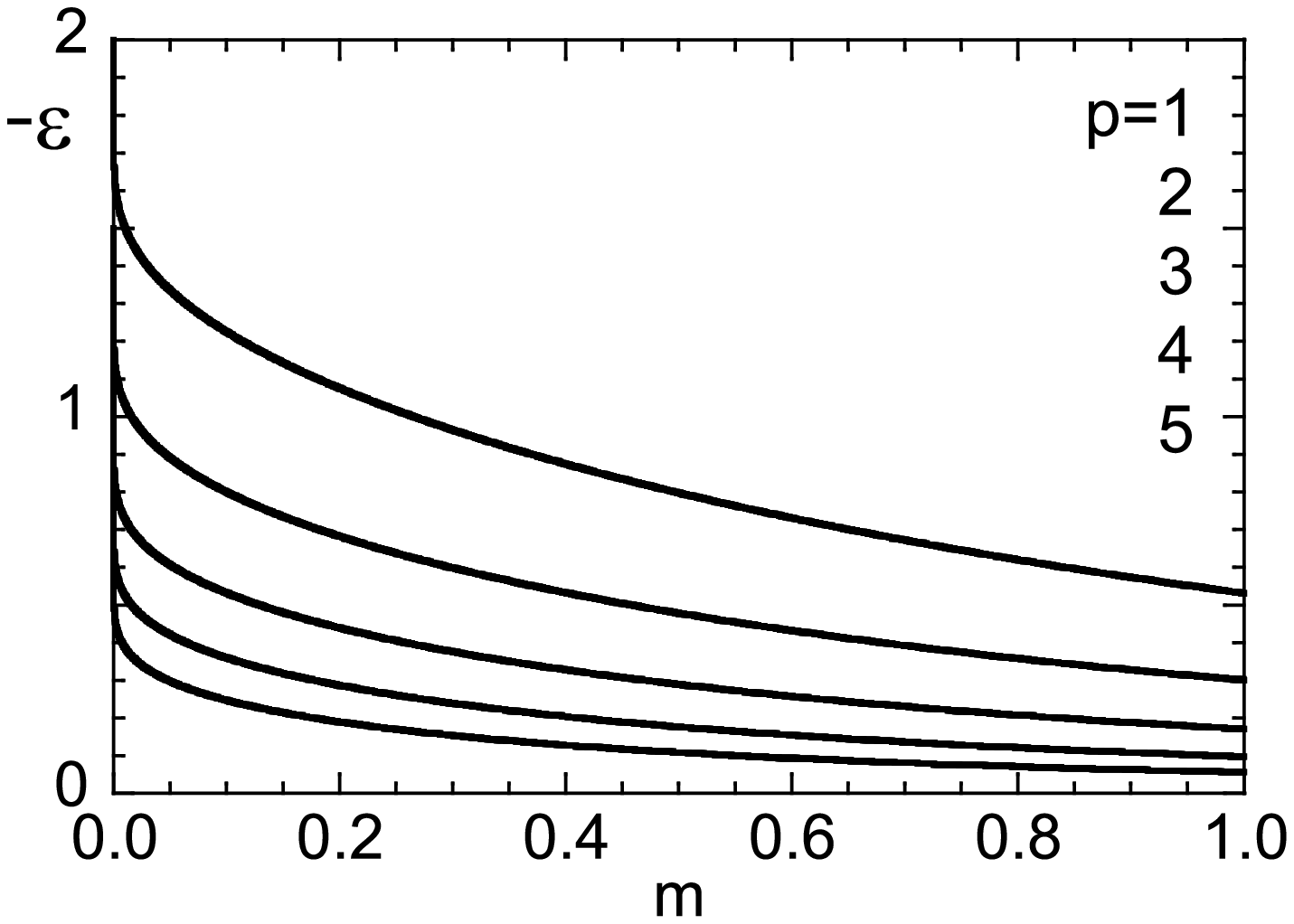}
\caption{Saddle-point equation (\ref{spmc}) 
in the microcanonical ensemble at $d=2$.
The function is finite at the origin.}
\label{sp-m2}
\end{center}
\end{minipage}
\end{figure}
\end{center}

 Let us first write the lattice Green function (\ref{Green}) as,  using 
 the variable $m^2=2(\tilde{z}-d)$,
\be
 G(k,\tilde{z}) 
 = \frac{2}{m^2+\sum_\mu\big(2\sin(k_\mu/2)\big)^2}.
\ee
 This is a decreasing function of $m$ and 
 the value at the origin $m=0$ determines the infrared behaviour.
 When $d=2$, $g(\tilde{z})$ defined in~(\ref{gz}) diverges logarithmically
 at $m=0~(\tilde{z}=2)$.

 We plot the right-hand sides of (\ref{spc}) and  (\ref{spmc}) in figures~\ref{sp-c2}
 and \ref{sp-m2}, respectively, for several values of $p$.
 In the canonical case, the right-hand side of~(\ref{spc})
 diverges at $\tilde{z}=d=2$ ($m=0$) and 
 is monotonically decreasing when $p\le 4$.
 Therefore, in this case, for a given $\beta$, $\tilde{z}$
 is determined uniquely.
 When $p>4$, in a certain range of $\beta$, 
 three solutions exist and  
 $m$ is not determined uniquely.
 This is understood as the emergence of a
 first-order transition~\cite{CP}.
 On the other hand, the function in~(\ref{spmc}) 
 for the microcanonical ensemble 
 is finite at the origin ($m=0$) and 
 is a decreasing function for arbitrary $p$. 
 Since the functional value at the origin 
 corresponds to the ground-state energy, 
 the solution is determined uniquely 
 for a given $\epsilon$ larger than or equal to the ground-state energy.
 Nothing singular happens in this case.

\begin{center}
\begin{figure}[htb]
\begin{minipage}[h]{0.5\textwidth}
\begin{center}
\includegraphics[width=0.9\columnwidth]{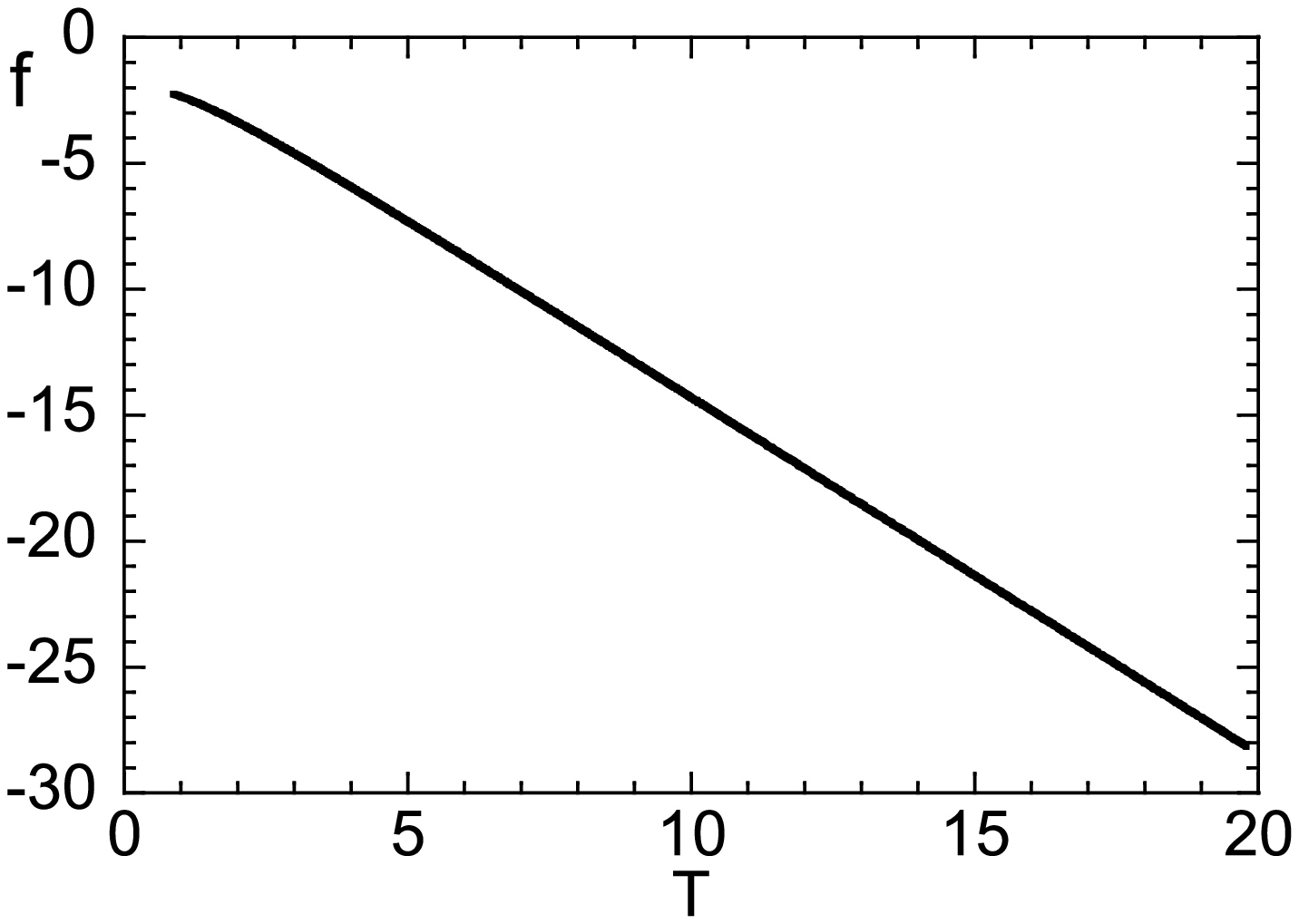}
\caption{Free energy density $f(T)$ 
in the canonical ensemble for $d=2$, $p=1$.}
\label{f1-c2}
\end{center}
\end{minipage}
\begin{minipage}[h]{0.49\textwidth}
\begin{center}
\includegraphics[width=0.9\columnwidth]{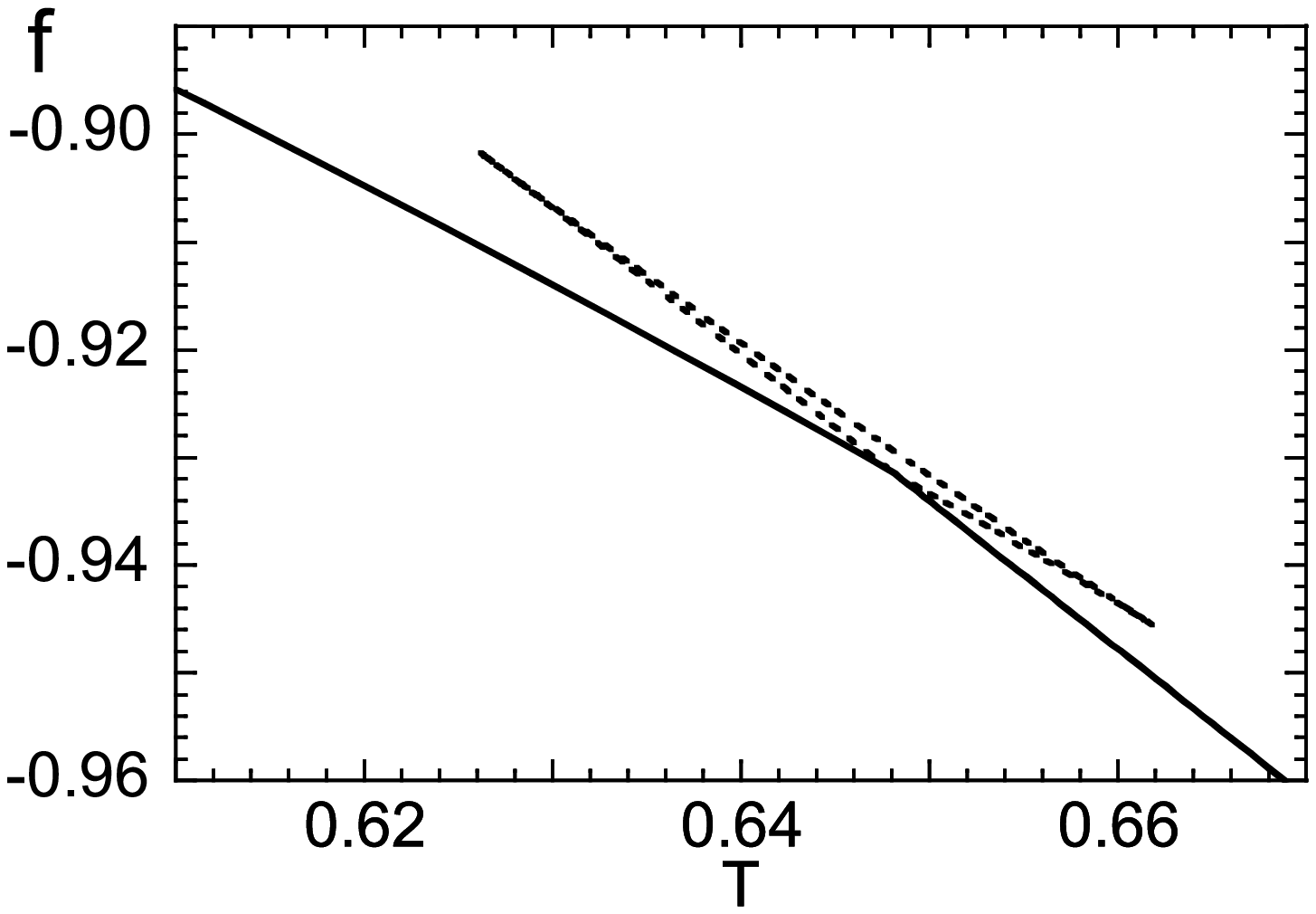}
\caption{$f(T)$ in the canonical ensemble for $d=2$, $p=5$.
The dotted line denotes the thermodynamically irrelevant
(unstable or metastable) saddle-point solutions.}
\label{f5-c2}
\end{center}
\end{minipage}
\end{figure}
\end{center}
\begin{center}
\begin{figure}[htb]
\begin{minipage}[h]{0.5\textwidth}
\begin{center}
\includegraphics[width=0.9\columnwidth]{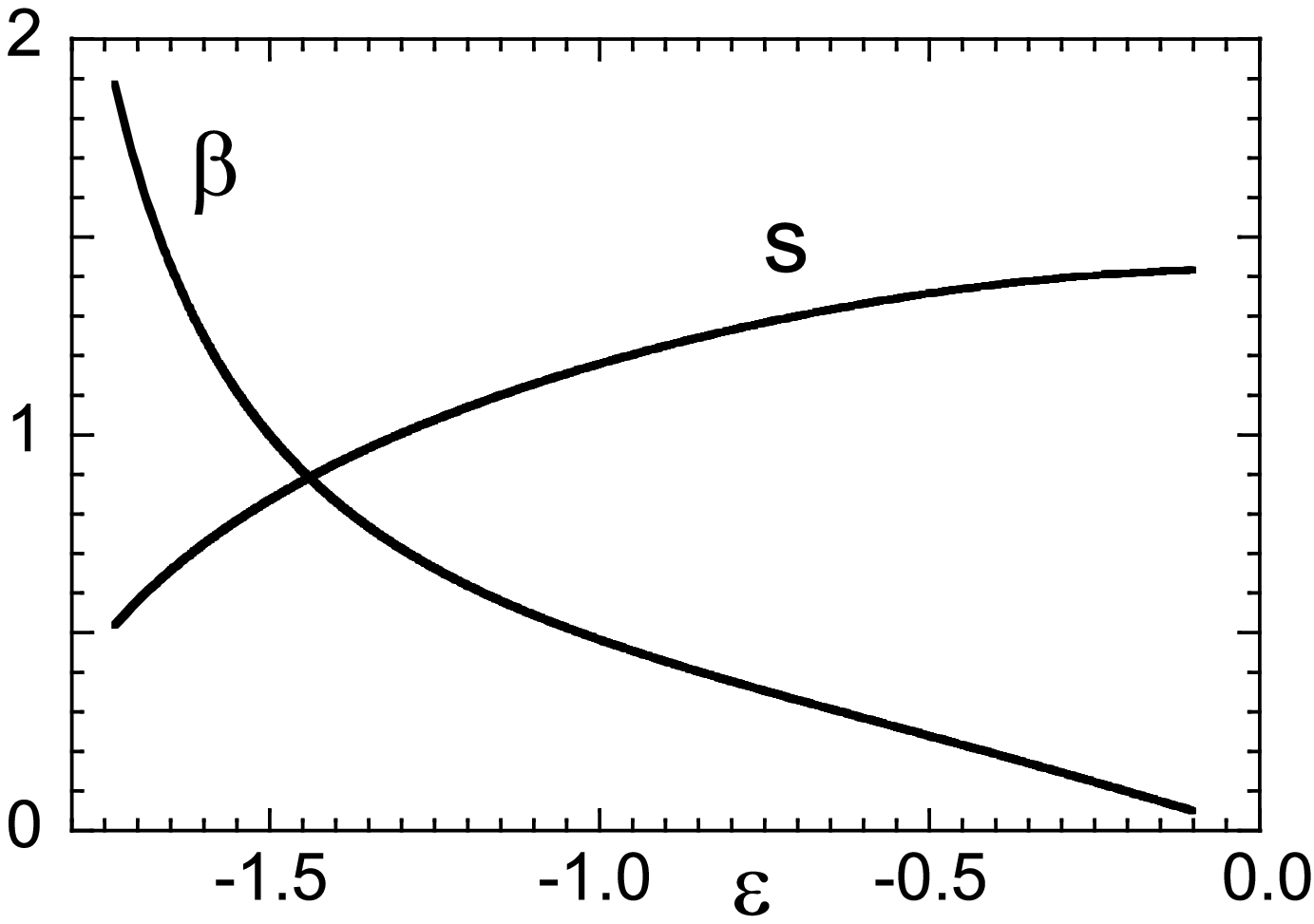}
\caption{Entropy density $s(\epsilon)$ and 
inverse temperature $\beta(\epsilon)=ds(\epsilon)/d\epsilon$
 in the microcanonical ensemble for $d=2$, $p=1$.}
\label{s1-m2}
\end{center}
\end{minipage}
\begin{minipage}[h]{0.49\textwidth}
\begin{center}
\includegraphics[width=0.9\columnwidth]{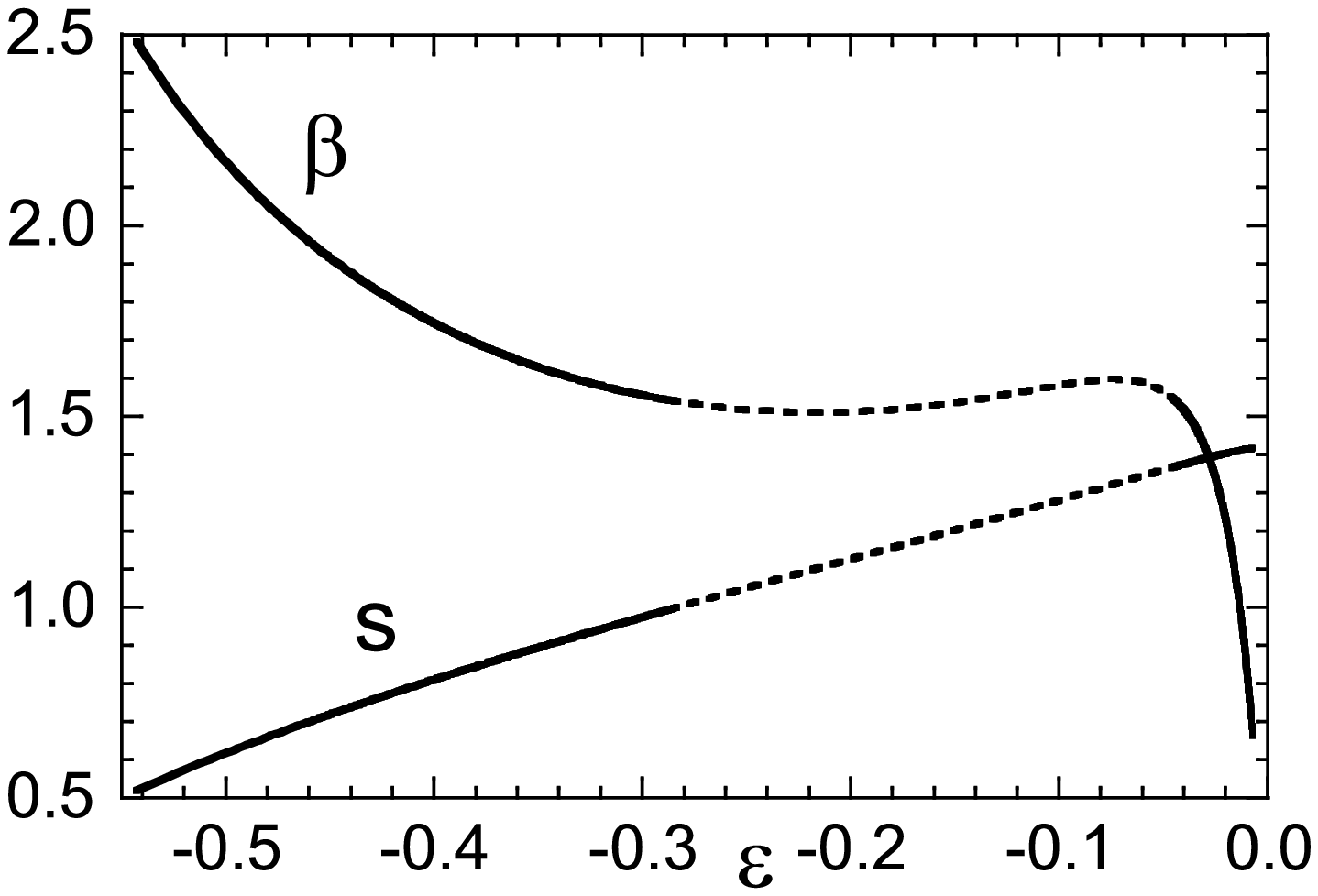}
\caption{$s(\epsilon)$ and $\beta(\epsilon)$
 in the microcanonical ensemble for $d=2$, $p=5$.
 $s(\epsilon)$ is non-concave and correspondingly 
 $\beta(\epsilon)$ is non-monotonic.
 The dotted parts correspond to the metastable and
 unstable branches in the canonical ensemble
 in figure \ref{f5-c2}.
}
\label{s5-m2}
\end{center}
\end{minipage}
\end{figure}
\end{center}

 From the obtained saddle-point solution in the canonical ensemble, 
 we plot the free-energy density $f$ for $p=1$ and 5 
 in figures~\ref{f1-c2} and \ref{f5-c2}, respectively.
 We see that a first-order transition occurs when $p=5$.
 We also plot the entropy density $s$ 
 and the inverse temperature $\beta=ds/d\epsilon$ 
 in the microcanonical ensemble
 for $p=1$ and 5 in figures~\ref{s1-m2} and \ref{s5-m2}, respectively.
 For $p=5$ in figure~\ref{s5-m2}, we see that 
 $\beta$ is non-monotonic (the entropy is non-concave) 
 for $-0.21<\epsilon<-0.07$  
 and consequently the specific heat is negative.
 In this sense, ensembles may seem inequivalent.
 This behaviour is similar to the case of systems with long-range
 interactions, where the mean-field picture applies.
 A remarkable fact is that this mean-field-like behaviour has been found by
 exact calculations for the two-dimensional system with short-range interactions.

\subsection{Phase separation}

 Ensemble equivalence is recovered in the present system if we choose 
 a proper saddle-point solution which represents a phase-separated state.
 We have assumed in~(\ref{s}) that the auxiliary variables are 
 independent of the subscripts $i$ and $j$.
 It implies that the phase is uniform in space.
 In order to describe the situation with phase separation,  
 we divide the system into two parts with
 $N_1$ and $N_2$ spins, respectively.
 The particular shape of the two sub-regions is not important, as far
 as they are geometrically compact objects, with a surface-to-volume ratio
 that vanishes in the thermodynamic limit 
 (for instance, a cubic lattice may be divided into two slabs).
 We set auxiliary variables in each subsystem as
 $z^{(1)},\rho^{(1)},\lambda^{(1)}$ and 
 $z^{(2)},\rho^{(2)},\lambda^{(2)}$, respectively.
 In the thermodynamic limit, we expect that the interface terms between 
 two subsystems are irrelevant due to the short-range nature of the system.
 We prove it rigorously in~\ref{app:bound}.
 Then, the number of states $\Omega$ is written as the sum of 
 contributions from two subsystems as 
\be
 \Omega (\epsilon)&=& \int\frac{dt}{2\pi}\exp\Biggl[
  iNnJt\left\{\epsilon
 +\frac{N_1}{N}dV(\rho^{(1)})
  +\frac{N_2}{N}dV(\rho^{(2)})\right\} 
 \no\\
 & & 
  +Nn\left(\frac{N_1}{N}z^{(1)}+\frac{N_2}{N}z^{(2)}\right)
 -Nn\left(
  \frac{N_1}{N}d\lambda^{(1)}\rho^{(1)}
  +\frac{N_2}{N}d\lambda^{(2)}\rho^{(2)}
 \right)
 \no\\
 & & 
  +n\ln\Tr\exp\Biggl(
 -\sum_{i=1}^{N_1}z^{(1)}(S_i^{(1)})^{2}
  +\sum_{\langle ij\rangle}\lambda^{(1)}S_i^{(1)}S_j^{(1)}
 \Biggr)
 \no\\
  & & 
 +n\ln\Tr\exp\Biggl(
 -\sum_{i=1}^{N_2}z^{(2)}(S_i^{(2)})^2
  +\sum_{\langle ij\rangle}\lambda^{(2)}S_i^{(2)}S_j^{(2)}
 \Biggr)\Biggr],
\ee
 where $S_i^{(1)}$ ($S_i^{(2)}$) 
 represents the spin variable in subsystem 1 (2).
 The saddle-point equations are written as 
 \be
 \label{e2} & & \epsilon 
 = \frac{N_1}{N}\epsilon_1+\frac{N_2}{N}\epsilon_2, \\
 & & \epsilon_{i} = -dV\left(\frac{\tilde{z}^{(i)}}{d}
 -\frac{1}{dg(\tilde{z}^{(i)})}\right) \quad (i=1, 2),
\ee
 where $\tilde{z}_1=z^{(1)}/\lambda^{(1)}$ and
 $\tilde{z}_2=z^{(2)}/\lambda^{(2)}$.
 Then, the entropy density is expressed as 
\be
 \label{s2} & & s = \frac{N_1}{N}s_1+\frac{N_2}{N}s_2, \\
 & & s_{i} =
 -\frac{1}{2}\ln g(\tilde{z}^{(i)})
 -\frac{1}{2}\int \frac{d^dk}{(2\pi)^d}\ln
 \left(\tilde{z}^{(i)}-\sum_{\mu=1}^d\cos k_\mu\right)
 +\frac{1}{2}
 +\frac{1}{2}\ln 2\pi. 
\ee

\begin{center}
\begin{figure}[htb]
\begin{center}
\includegraphics[width=0.6\columnwidth]{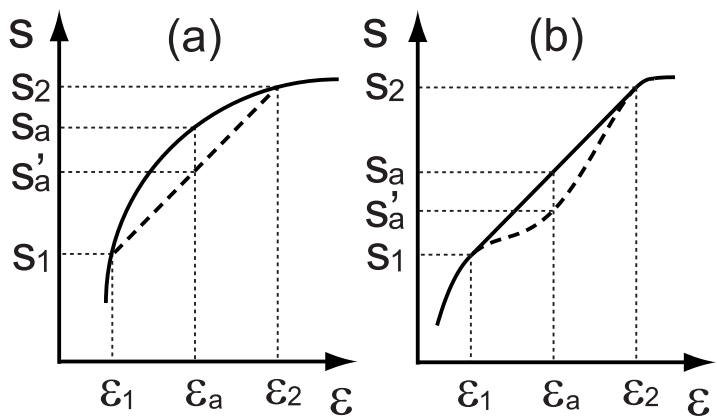}
\caption{
Comparison of the entropies
obtained from the uniform and phase-separated solutions.
The entropy from the phase-separated solution is represented 
by the straight lines.
(a) When the entropy for the uniform solution is concave, 
the phase-separated solution is irrelevant because it has a smaller value
of the entropy $s'_a$ than the true $s_a$.
(b) When the entropy is non-concave, 
a first-order transition occurs 
and a phase-separated state is realized
between $\epsilon_1$ and $\epsilon_2$ because it has a larger entropy.
}
\label{se}
\end{center}
\end{figure}
\end{center}

 A notable fact is that these expressions hold for 
 any ratio of the separated phases, $c=N_1/N, 1-c=N_2/N$
 as well as for any choice of $\epsilon_{1,2}$ and $s_{1,2}$.
 Thus, we should discuss what values of these parameters are actually chosen 
 for a given fixed value of the energy $\epsilon=\epsilon_a$.
 First, if the total entropy is concave, the hypothetical phase separation
 means that the value of the entropy would be $s'_a=c s_1 +(1-c)s_2$,
 which is lower than the true entropy $s_a$
 as can be understood from figure~\ref{se}(a).
 Thus, there is no phase separation in the stable state.
 Technically, this means that the exponent of the integral 
 for the number of states $\Omega =e^S$ becomes largest 
 at the saddle point representing the uniform state,
 not at the point corresponding to the phase-separated state.

 On the other hand, the situation is different
 when the total entropy is non-concave as we show in figure~\ref{se}(b).
 At $\epsilon=\epsilon_a$, 
 the state with the entropy $s=s_a=c s_1 +(1-c)s_2$ 
 is more stable than that with $s'_a$ and is realized as the
 phase-separated state in the usual sense.
 Technically, the saddle point corresponding to this former state has the
 largest contribution to the integral.
 Thus, we can obtain the phase-separated state 
 by relaxing the uniformity condition of the saddle-point solution.
 It should be noticed that only the microcanonical solution
 needs this non-uniform prescription of the saddle-point values.
 The uniform solution for the canonical case 
 (\ref{saddle_point_eq_canonical}) shows no inconsistencies.

\subsection{$d=3$}

\begin{center}
\begin{figure}[htb]
\begin{minipage}[h]{0.5\textwidth}
\begin{center}
\includegraphics[width=0.9\columnwidth]{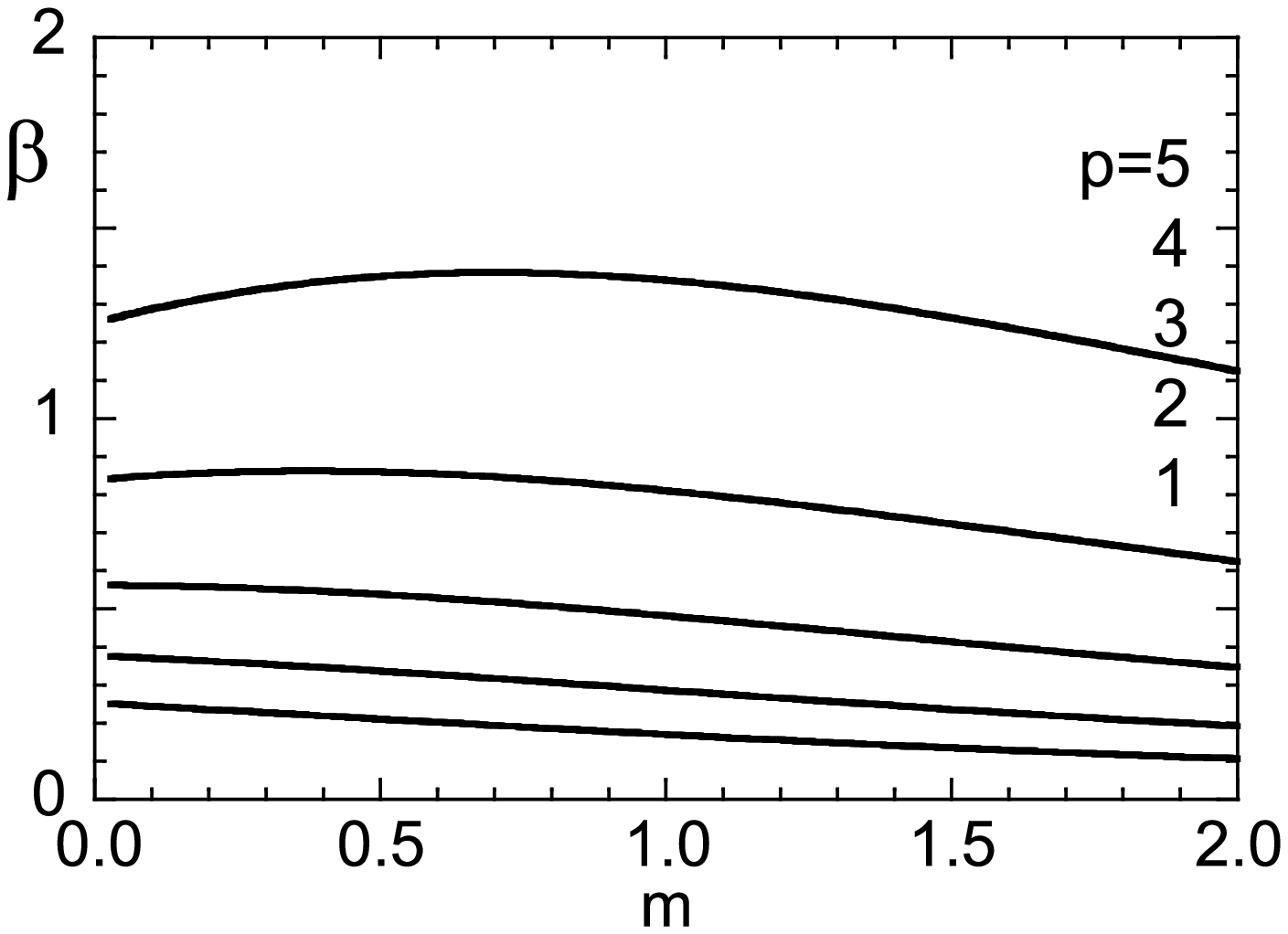}
\caption{Saddle-point equation (\ref{spc})
 in the canonical ensemble at $d=3$.}
\label{sp-c3}
\end{center}
\end{minipage}
\begin{minipage}[h]{0.5\textwidth}
\begin{center}
\includegraphics[width=0.9\columnwidth]{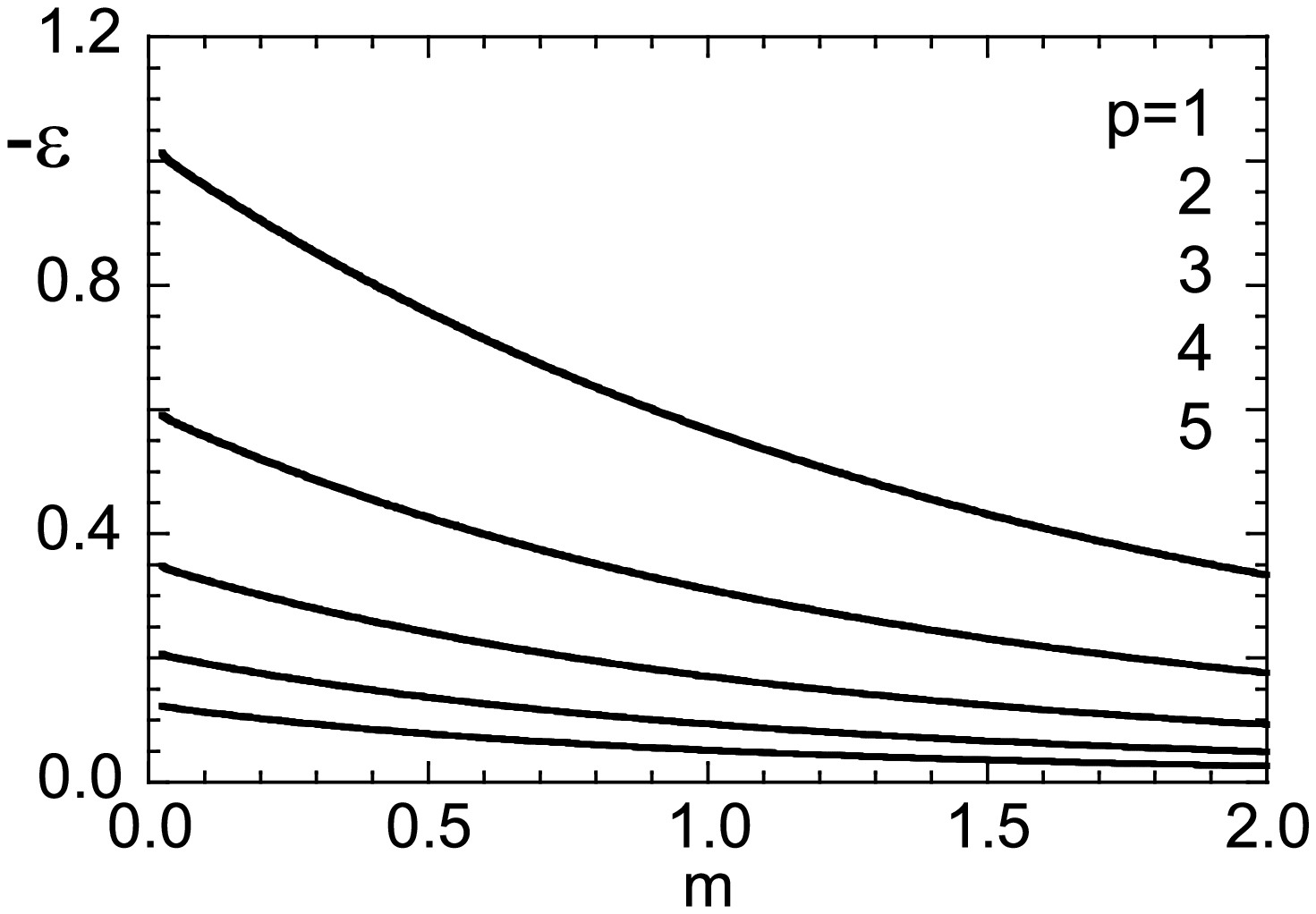}
\caption{Saddle-point equation (\ref{spmc}) 
 in the microcanonical ensemble at $d=3$.
The value at the origin is not equal to the ground state energy.}
\label{sp-m3}
\end{center}
\end{minipage}
\end{figure}
\end{center}
\begin{center}
\begin{figure}[htb]
\begin{minipage}[h]{0.5\textwidth}
\begin{center}
\includegraphics[width=0.9\columnwidth]{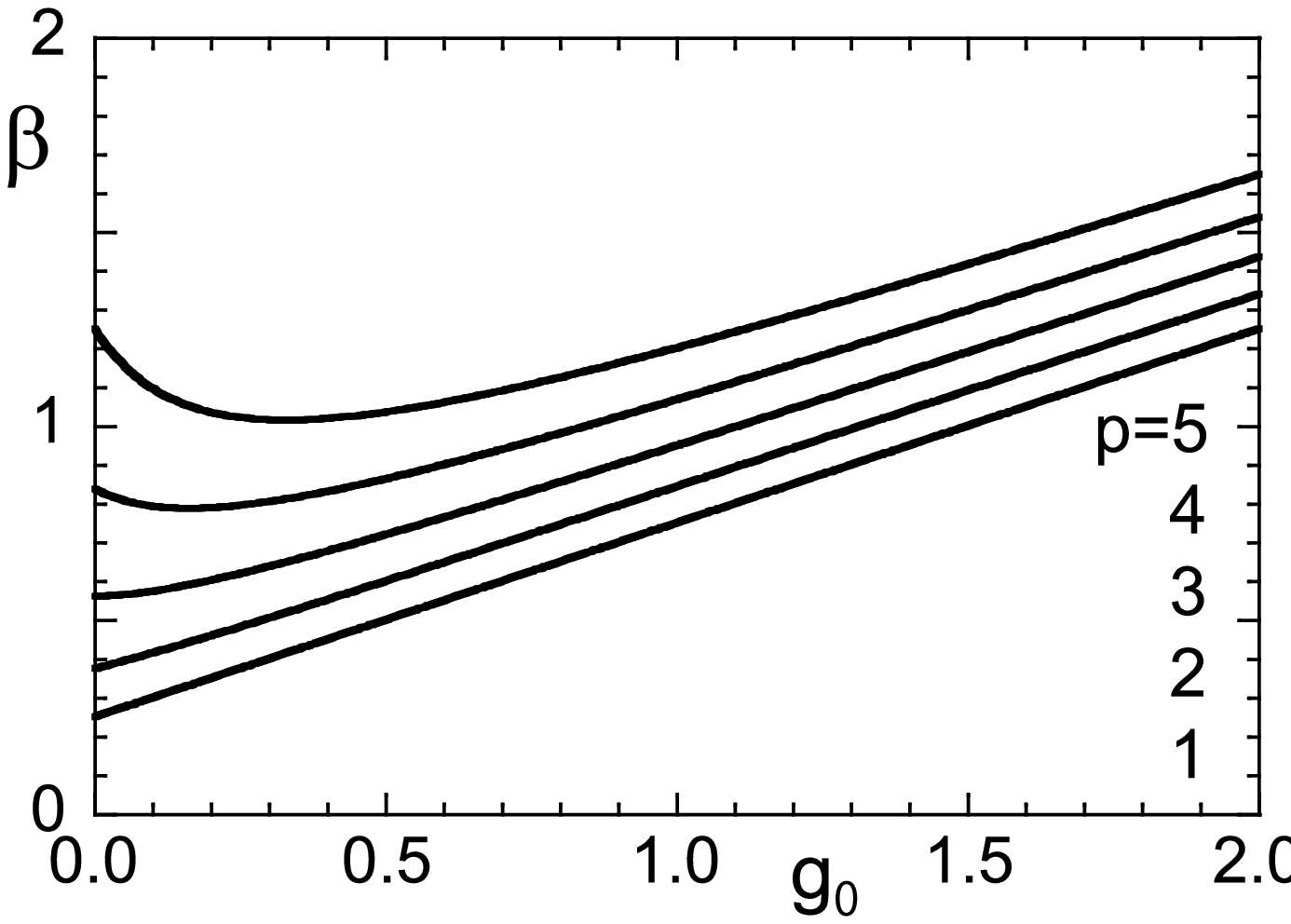}
\caption{Zero-mode part of the saddle-point equation 
 in the canonical ensemble at $d=3$.}
\label{g0-c3}
\end{center}
\end{minipage}
\begin{minipage}[h]{0.5\textwidth}
\begin{center}
\includegraphics[width=0.9\columnwidth]{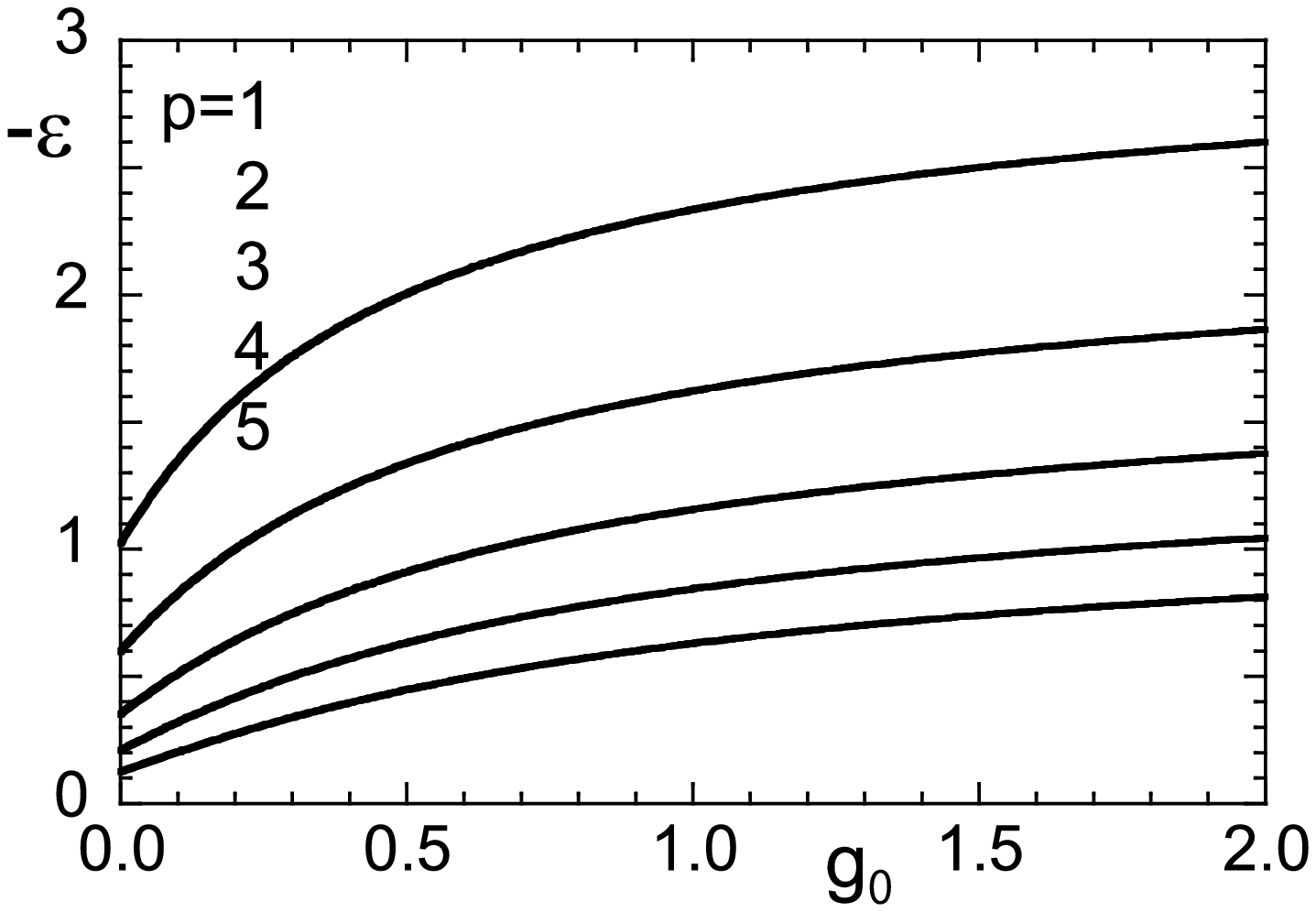}
\caption{Zero-mode part of the saddle-point equation 
 in the microcanonical ensemble at $d=3$.}
\label{g0-m3}
\end{center}
\end{minipage}
\end{figure}
\end{center}

 Let us next consider the three-dimensional system, in which case
 $g(\tilde{z})$ is finite at $m=0$ 
 ($g(\tilde{z}=3)\approx 0.505$) and is monotonically decreasing.
 As shown in figures~\ref{sp-c3} and \ref{sp-m3},
 the saddle-point equation has no solution 
 at low temperature or low energy.
 To avoid this difficulty, the zero mode $k=0$ in (\ref{gz})
 should be separated from the integral, 
 similarly to the Bose-Einstein condensation, as 
\be
 g(\tilde{z}) 
 \to \tilde{g}(k=0,\tilde{z})+g(\tilde{z})
 = \frac{1}{N}\frac{1}{\tilde{z}-3}
 +\int \frac{d^3k}{(2\pi)^3}\frac{1}{\tilde{z}-\sum_{\mu=1}^3\cos k_\mu}.
\ee
 The parameter $\tilde{z}$ approaches $3$ in the thermodynamic limit so that 
 the first term gives a finite contribution $g_0$ in this limit.
 Then, we can find the solution of the saddle-point equation  by the replacement 
\be
 g(\tilde{z})\to g_0+g(3).
\ee
 As depicted in figures~\ref{g0-c3} and \ref{g0-m3}, $g_0$ can be fixed
 from the saddle-point equation for a given $\beta$ (or $\epsilon$) below (or above)
 the values achievable in figure \ref{sp-c3} (or figure \ref{sp-m3}).
 Hence, there exists a solution for any $\beta$ or $\epsilon$, the latter
 being larger than or equal to the ground-state energy.

\begin{center}
\begin{figure}[htb]
\begin{minipage}[h]{0.5\textwidth}
\begin{center}
\includegraphics[width=0.9\columnwidth]{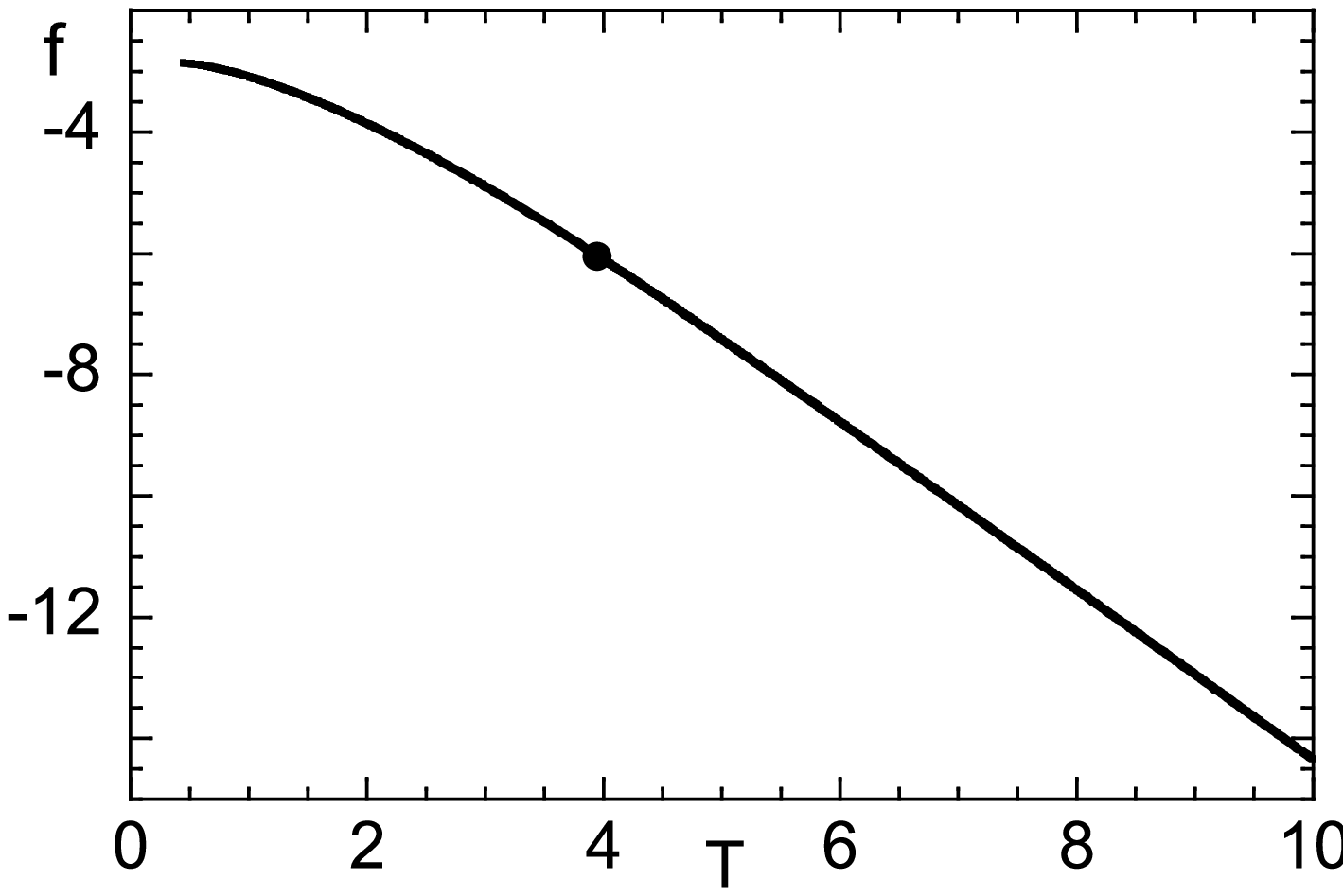}
\caption{$f(T)$ in the canonical ensemble for $d=3$, $p=1$.
 The dot denotes the transition point, where $g_0$ starts to be finite.}
\label{f1-c3}
\end{center}
\end{minipage}
\begin{minipage}[h]{0.49\textwidth}
\begin{center}
\includegraphics[width=0.9\columnwidth]{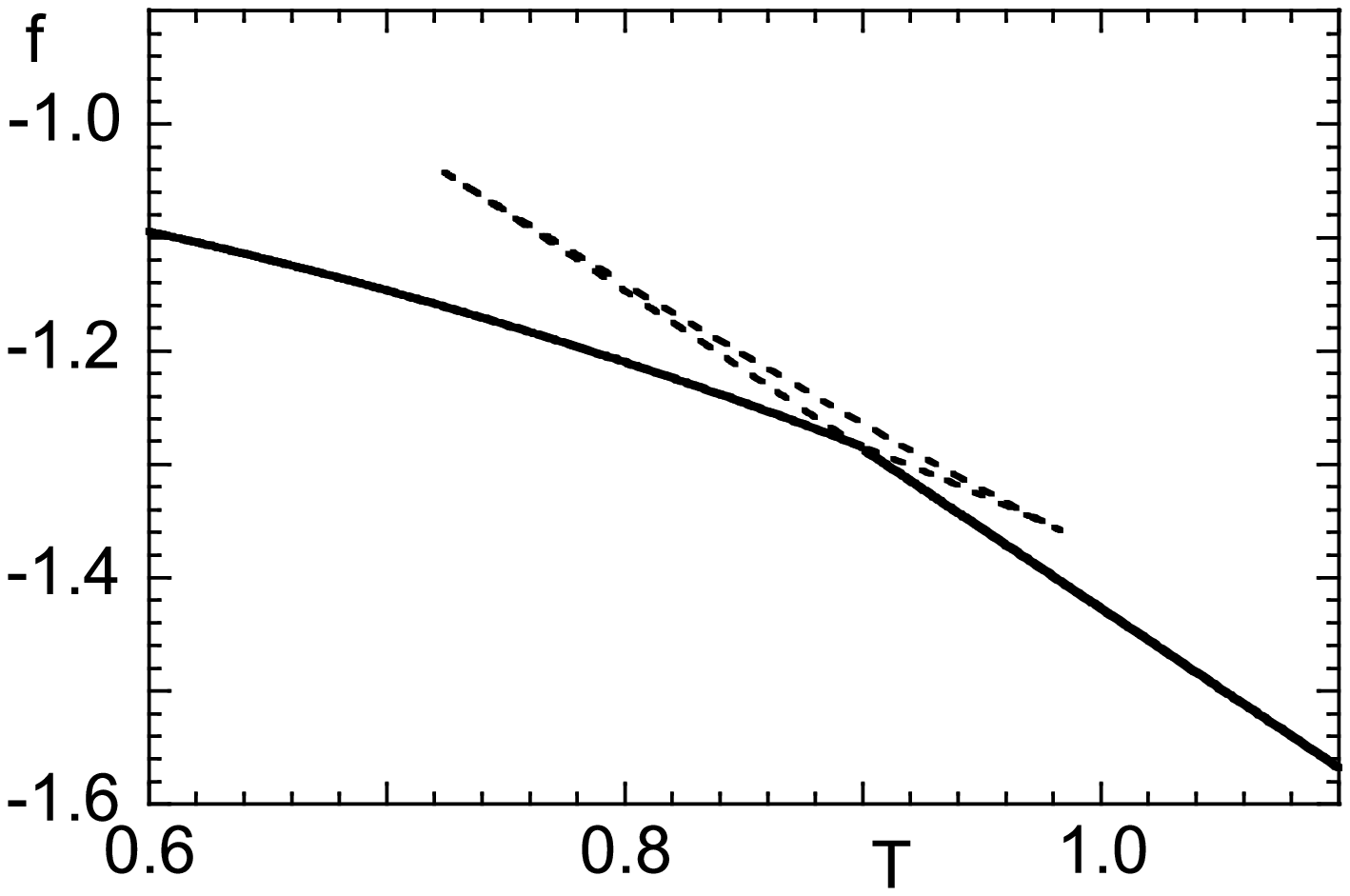}
\caption{$f(T)$ in the canonical ensemble for $d=3$, $p=5$.}
\label{f5-c3}
\end{center}
\end{minipage}
\end{figure}
\end{center}
\begin{center}
\begin{figure}[htb]
\begin{minipage}[h]{0.5\textwidth}
\begin{center}
\includegraphics[width=0.9\columnwidth]{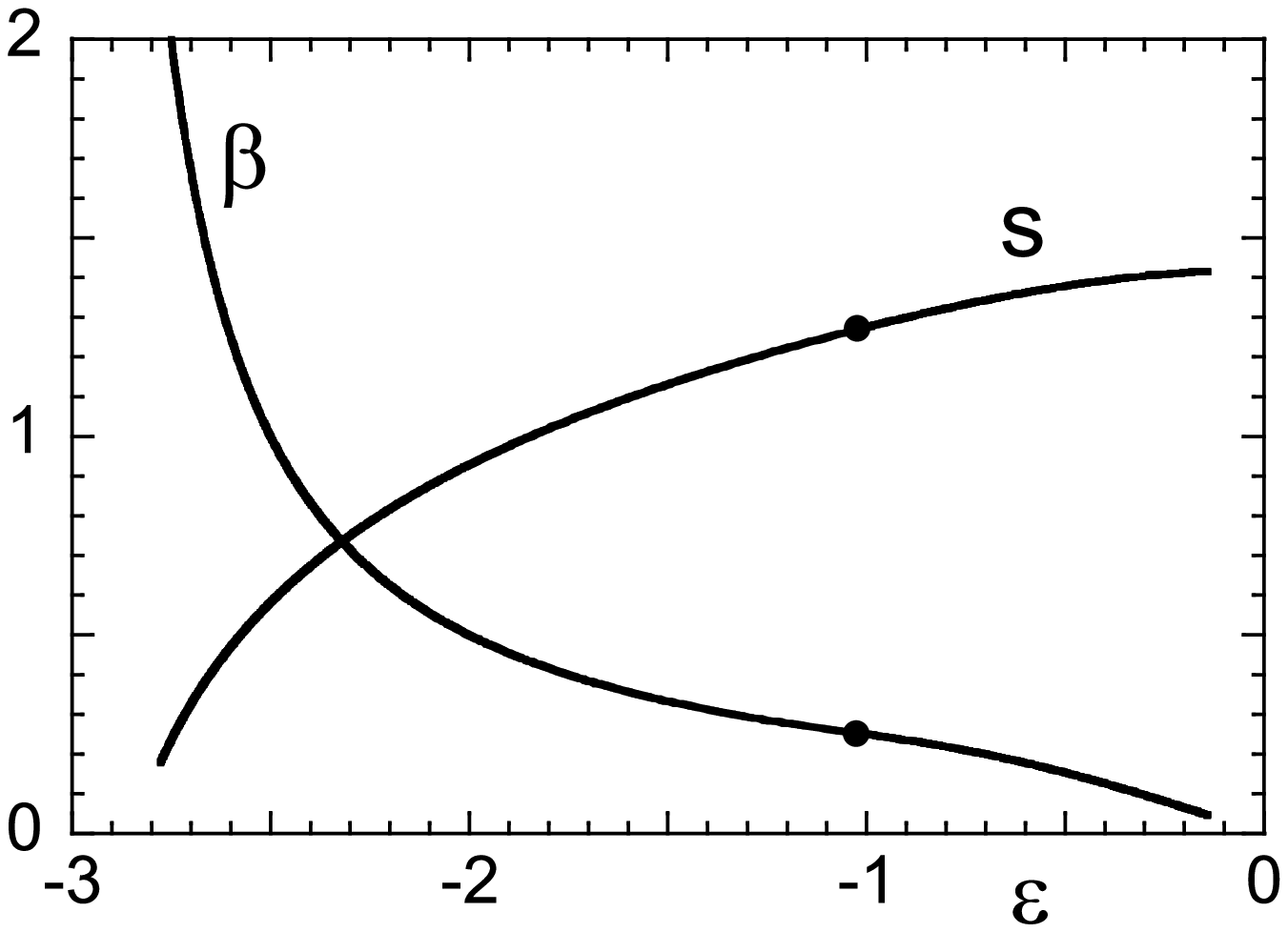}
\caption{$s(\epsilon)$ and $\beta(\epsilon)$
 in the microcanonical ensemble for $d=3$, $p=1$.
 The dot denotes the transition point.}
\label{s1-m3}
\end{center}
\end{minipage}
\begin{minipage}[h]{0.49\textwidth}
\begin{center}
\includegraphics[width=0.9\columnwidth]{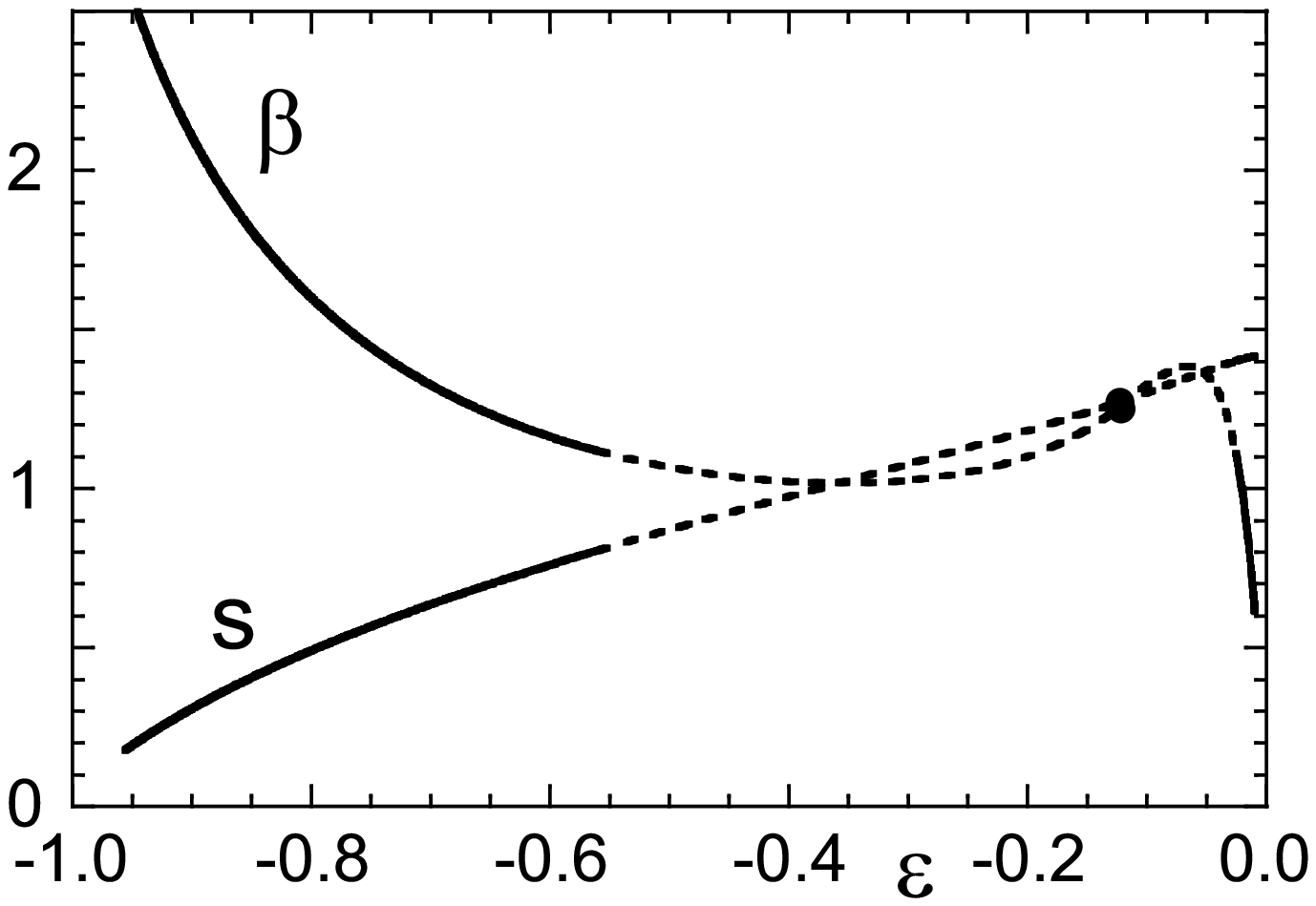}
\caption{$s(\epsilon)$ and $\beta(\epsilon)$
 in the microcanonical ensemble for $d=3$, $p=5$.
 The putative continuous transition point denoted by the dot
 is hidden by a first-order transition.
 The dotted parts correspond to those in figure \ref{f5-c3}.
}
\label{s5-m3}
\end{center}
\end{minipage}
\end{figure}
\end{center}

 We plot the free energy in the canonical ensemble 
 in figures~\ref{f1-c3} and \ref{f5-c3} for $p=1$ and $p=5$, respectively.
 A continuous transition to the zero-mode condensed phase  
 is observed for $p=1$.
 It is replaced by a discontinuous transition for $p=5$.
 From the microcanonical analysis, 
 we plot $s$ and $\beta$ in 
 figures~\ref{s1-m3} and \ref{s5-m3} for $p=1$ and $p=5$, respectively.
 As can be understood from these figures, 
 transitions between the condensed and non-condensed phases
 exist in both ensembles in three dimensions.
 It is discontinuous for $p=5$.
 Similarly to the $d=2$ case, the apparent ensemble inequivalence (negative
 specific heat in the microcanonical ensemble) can be avoided by 
 the proper choice of the saddle-point solution.

\section{Monte Carlo analysis}
\label{montecarlo}

 In order to check if the results of the previous section is specific to the
 spherical model, we study the $n=2$ model, the $XY$ model, in two dimensions
 by Monte Carlo simulations.
 A canonical Monte Carlo analysis of the present model has already been
 carried out in~\cite{DSS}, and
 the microcanonical case has been done in~\cite{Ota}.
 However, these previous studies are not sufficient to clarify the problem
 of ensemble equivalence, and we analyze the same model here from
 our own point of view.
 We refer the reader to \cite{BG} for a related study, where 
 the ensemble equivalence was examined in a mean-field model 
 on random graphs. 
 
 The original $XY$ model with linear interactions in two dimensions 
 exhibits the Kosterlitz-Thouless transition~\cite{KT}.
 In~\cite{SR}, a detailed study is reported on
 how the transition can be changed in nonlinearly-interacting systems.
 Since our aim is not to study this topological transition, only
 the caloric curve has been calculated in our Monte Carlo analysis.
\begin{center}
\begin{figure}[htb]
\begin{minipage}[h]{0.5\textwidth}
\begin{center}
\includegraphics[width=0.9\columnwidth]{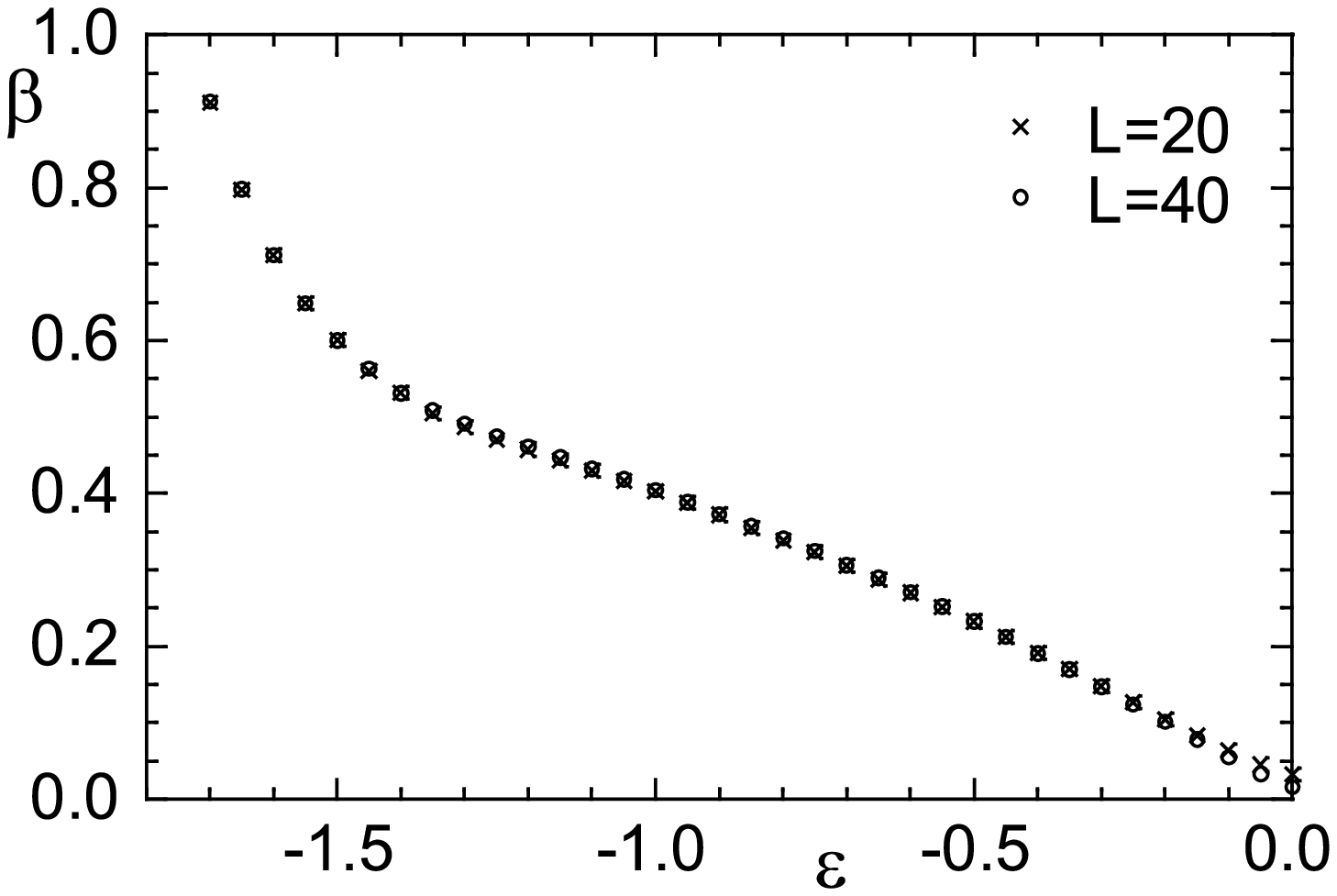}
\caption{$\beta(\epsilon)$ from
 the microcanonical Monte Carlo calculation 
 for $n=2$, $d=2$ and $p=1$.
 The error bars are smaller than the symbol size.}
\label{p01}
\end{center}
\end{minipage}
\begin{minipage}[h]{0.5\textwidth}
\begin{center}
\includegraphics[width=0.9\columnwidth]{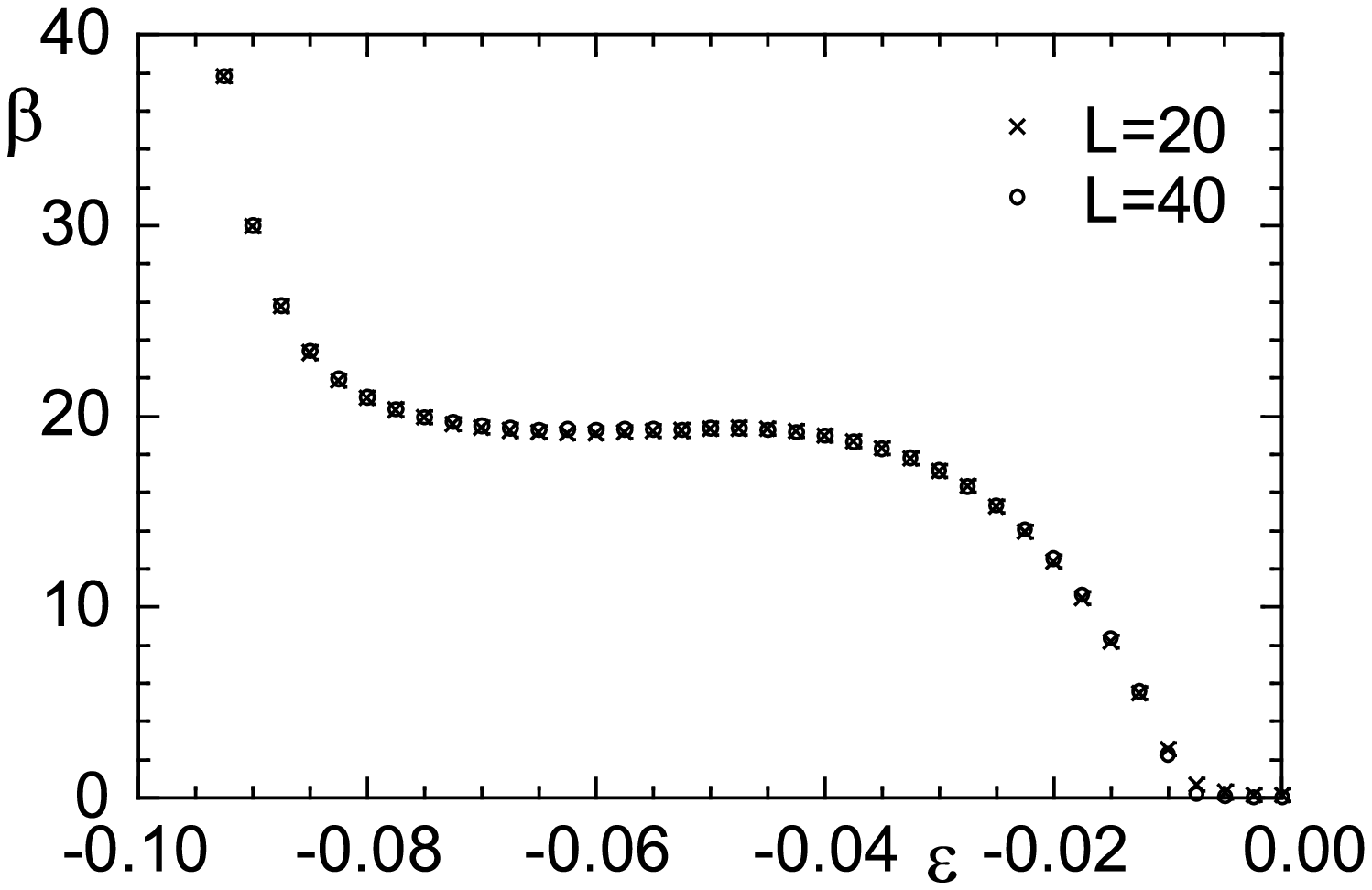}
\caption{$\beta(\epsilon)$  for $n=2$, $d=2$ and $p=40$.
The error bars are smaller than the symbol size.}
\label{p40}
\end{center}
\end{minipage}
\end{figure}
\end{center}
\begin{center}
\begin{figure}[htb]
\begin{minipage}[h]{0.5\textwidth}
\begin{center}
\includegraphics[width=0.9\columnwidth]{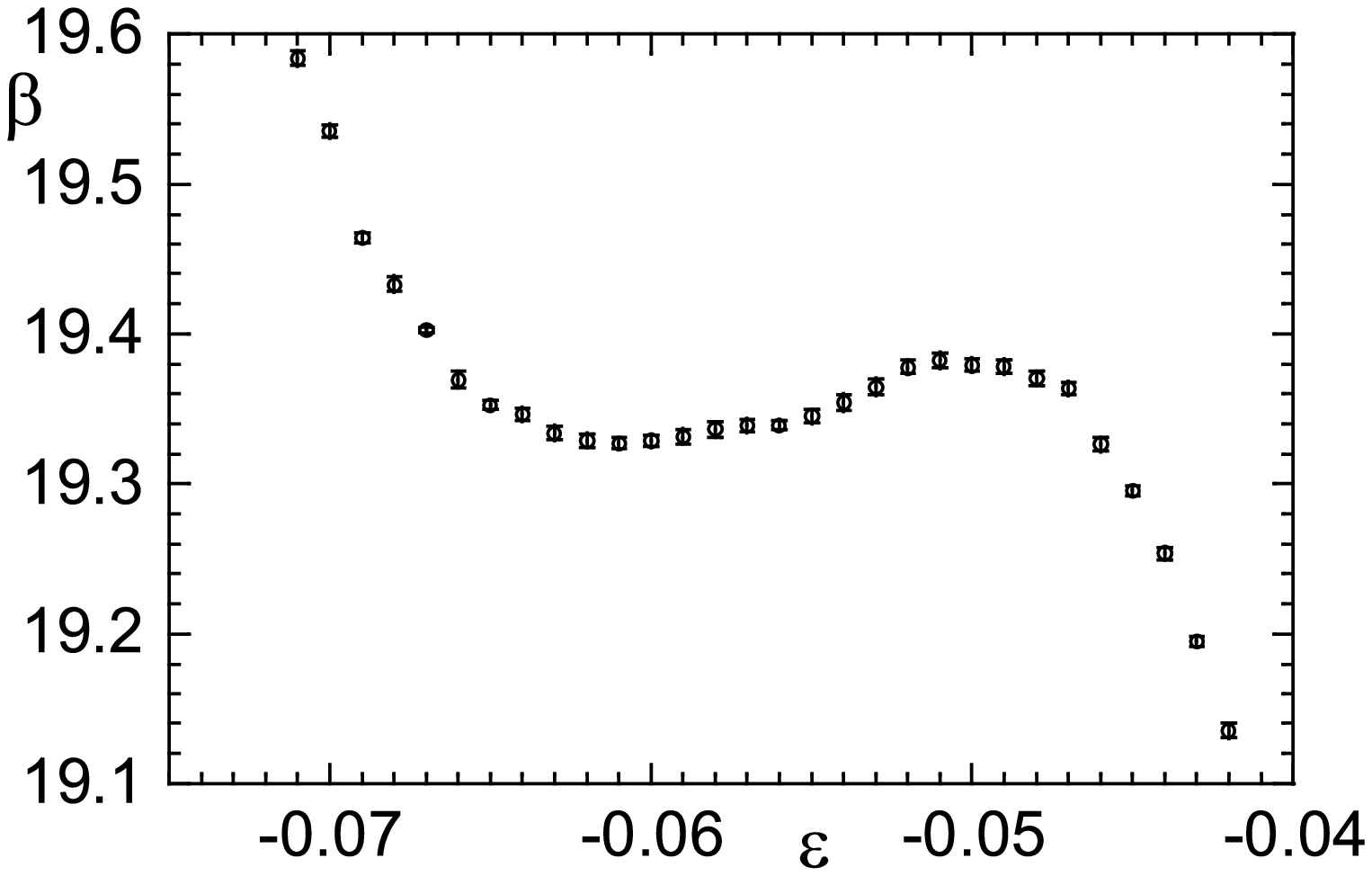}
\caption{$\beta(\epsilon)$ for $n=2$, $d=2$, $p=40$,
and $L=50$ near the non-monotonic region.
We have taken the average over 20 independent runs.
}
\label{p40l50}
\end{center}
\end{minipage}
\begin{minipage}[h]{0.5\textwidth}
\begin{center}
\includegraphics[width=0.9\columnwidth]{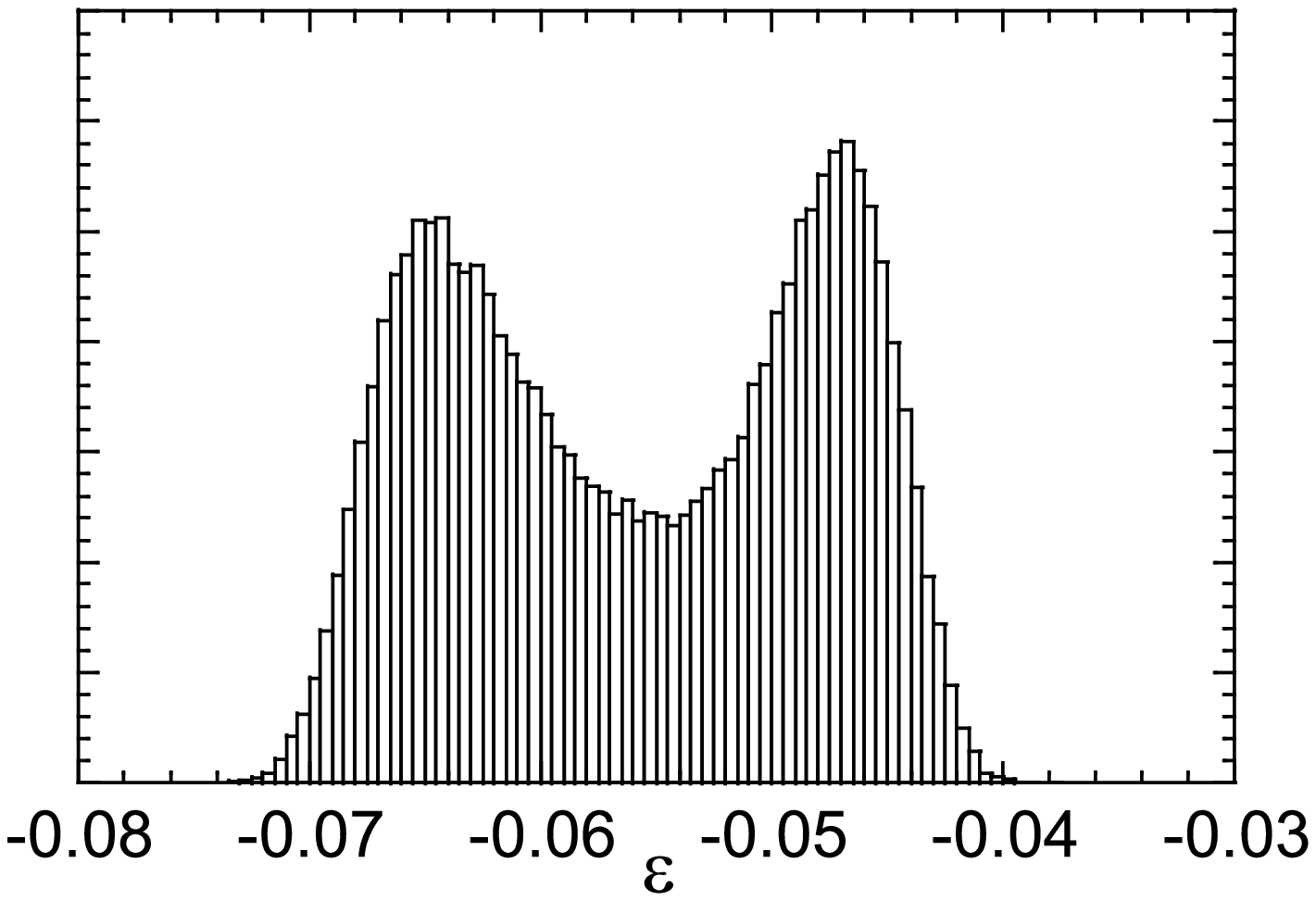}
\caption{Energy histogram in the phase coexistence region 
 from the canonical Monte Carlo calculation.
 We have chosen $n=2$, $d=2$, $p=40$, $L=50$, and $\beta=19.35$.
}
\label{p40l50c}
\end{center}
\end{minipage}
\end{figure}
\end{center}
  
 Several system sizes have been analyzed, $L=20$, 40, 50 and larger in some cases.
 To study the microcanonical ensemble, 
 we exploit the demon algorithm by Creutz~\cite{Creutz}.
 In this algorithm, for a given energy $\epsilon$, 
 the demon energy $E_{\rm D}$ is calculated 
 so that the sum of the system and demon energies is kept constant.
 Then, the inverse temperature is obtained from the probability 
 distribution ${\rm Prob}(E_{\rm D})\sim\exp(-\beta E_{\rm D})$.
 We have performed $10^6$ Monte Carlo steps per spin for each run.

 In figures~\ref{p01} and \ref{p40}, 
 we plot the energy dependence of the inverse temperature 
 for $p=1$ and $p=40$, respectively.
 For $p=1$, we see that 
 $\beta$ is a monotonically decreasing function of $\epsilon$.
 It is different for $p=40$, where the function shows a non-monotonic behaviour.
 Figure~\ref{p40l50} highlights this property for $L=50$.
 For a given $\beta$, $\epsilon$ is not determined uniquely 
 in a narrow region, which suggests the existence of negative specific heat.
 We have performed
 canonical Monte Carlo calculations using the simple Metropolis algorithm 
 to see the energy histogram, and the result is depicted
 in figure~\ref{p40l50c}, which clearly shows
 that a first-order transition exists in the form of phase coexistence.
 We have confirmed that the non-monotonic region of the caloric curve
 remains up to the size $L=100$ in the microcanonical simulations.
 An extrapolation suggests that it would persist to the thermodynamic limit.
 Thus, the negative specific heat seems to exist
 in the microcanonical ensemble also in the two-component system 
 as in the spherical model.

 We speculate that a reason for the apparent ensemble inequivalence for $p=40$ in
 Monte Carlo simulations may be that a phase separation, as discussed above for the
 spherical model, has not been realized in our simulations because of a very long
 relaxation time:
 The system has to spontaneously break up into two spatially separated regions
 with different macroscopic states, which could take a very long time to be realized
 in the microcanonical simulations.

\section{Summary and conclusion}
\label{conclusion}

 We have studied the $n$-vector model ($O(n)$-symmetric model) with nonlinear
 short-range interactions in two and three dimensions.
 The exact solution of the spherical model shows ostensible inequivalence of
 canonical and microcanonical ensembles through negative specific heat in the
 latter ensemble.
 We have argued that this paradox can be resolved  by explicitly taking into
 account a phase separation,
 which increases the entropy (thus increases the thermodynamic stability)
 in the microcanonical ensemble.
 It is noticed that the proper choice of the saddle-point solution 
 is required in the microcanonical ensemble to represent 
 the state with phase separation.
 Such a solution must be considered
 when the uniform ansatz of the
 saddle-point solution gives a non-concave entropy.
 We note that this procedure is needed only in the microcanonical ensemble. 
 In the canonical ensemble, 
 the uniform solution is sufficient to represent the stable state of the system.
 Another interesting aspect is that the exact solution of the spherical model
 is similar to the mean-field solutions applicable to long-range interacting models
 in the sense that the calculations of the  (canonical) partition function
 and the (microcanonical) entropy reduced to steepest descent integrations.
 The difference between the short- and long-range systems 
 is apparent when we consider the phase-separated state. 
 The interface term cannot be neglected in the long-range system, 
 which can be understood as an important source of ensemble inequivalence.
 Our analysis has succeeded to highlight this difference 
 in an exactly solvable  example.
 
 The $XY$ model with nonlinear interactions has been shown to behave
 similarly by Monte Carlo simulations in two dimensions.
 Discrepancies between ensembles may in this case be due to the
 long relaxation time to the fully-stable phase-separated state
 in the microcanonical simulation.
 We expect that a more elaborate method such as 
 the one developed in~\cite{MM} may resolve this problem.

 The $n$-vector models with nonlinear short-range interactions have been known
 to have the unusual property of the existence of first-order phase
 transitions even in two (and higher) dimensions~\cite{DSS}-\cite{vES2}.
 We have identified an additional highly non-trivial property of
 apparent ensemble inequivalence, which we expect to
 stimulate further studies of this very unusual class of models.

\appendix
\section{Derivation of the entropy in the microcanonical ensemble}
\label{app:smicro}

 We consider the number of states for a given energy $E$ 
\be
 \Omega = \Tr \delta(E-H) = \int\frac{dt}{2\pi}
 \Tr e^{i(E-H)t}.
\ee
 This expression has a similar form to the partition function 
 in the canonical ensemble except for the integral 
 over $t$ and the factor $e^{iEt}$.
 We may thus replace $\beta$ in the partition function by $it$.
 Therefore, the calculation goes along the same line 
 as in the canonical case and we can write 
\be
 \Omega 
 &=&  \int\frac{dt}{2\pi} \int\prod_{i}dz_i\prod_{\langle ij\rangle}
 d\lambda_{ij} d\rho_{ij}\, \exp\left[
 it\Biggl(E+Jn\sum_{\langle ij\rangle}V\left(\rho_{ij}\right)\Biggr)
 +n\sum_{i=1}^Nz_i
 \right.\no\\
 & & \left.
 -n\sum_{\langle ij\rangle}\lambda_{ij}\rho_{ij}
 +n\ln\Tr\exp\left(
 -\sum_{i=1}^Nz_iS_i^2+\sum_{\langle ij\rangle}\lambda_{ij}S_iS_j
 \right)\right].
\ee
 Then, we impose the uniform ansatz for 
 $z_i$, $\rho_{ij}$ and $\lambda_{ij}$ and
 obtain the number of states as
\be
 \Omega &=&  \exp\Bigg[
 it\Bigl(E+NndJV(\rho)\Bigr)
 +Nnz-Nnd\lambda\rho
 \no\\
 & & +n\ln\Tr\exp\Biggl(
 -\sum_{i=1}^N zS_i^2+\sum_{\langle ij\rangle}\lambda S_iS_j\Biggr)
 \Biggr], 
\ee
 and the saddle-point conditions
\be
 & & \epsilon+dV(\rho) = 0, \quad
 \lambda = it JV'(\rho), \quad
 2\lambda = g(\tilde{z}), \quad
 d\rho = \tilde{z}-\frac{1}{2\lambda},
\ee
 where $\epsilon=E/NnJ$.
 Combining these results, we finally obtain (\ref{spmc}) and (\ref{Omega}).

\section{Bounds on interface effects in the free energy of the Gaussian model}
\label{app:bound}

 In order to rigorously justify the calculations using only 
 the two phase-separated regions without interface terms, 
 let us estimate the order of magnitude of the effects
 that the interface terms have on the free energy of the Gaussian model.
 We define the free energy of the Gaussian model as
\be
 h(x)=\ln \Tr\exp\left(-z\sum_i S_i^2+\lambda 
 \sum_{\langle ij \rangle}^{(1)}S_i S_j
 +\lambda x\sum_{\langle ij \rangle}^{(2)}S_i S_j\right),
 \label{free-energy-h}
\ee
 where the summation with superscript (1) runs over all interactions
 within the two independent (phase-separated) subsystems and 
 the summation with superscript (2) is 
 for interactions across the interface.
 The parameter $\lambda$ will be assumed to be positive
 without losing generality on a bipartite lattice.
 Notice that we have assumed that the interactions have common values
 in the two subsystems.
 We will show later that this restriction can be removed.
 The boundary conditions are assumed to be free in the $x$ direction
 and periodic otherwise.
 Here the term `interface' stands for the region 
 in the middle of the system that runs perpendicular to 
 the $x$ axis and separates two subsystems,
 whereas the `boundary' is for the outmost sites of the total system.
 Equation (\ref{free-energy-h}) indicates that 
 the interface interactions have the strength $\lambda x$ and 
 all other interactions have $\lambda$.

 Our goal is to prove that 
\be
 |h(1)-h(0)|\le c N_b, \label{ineq1}
\ee
 where $N_b$ is the number of interactions across the interface and
 $c$ is a quantity asymptotically independent of $N_b$ and $N$ 
 (total number of sites).
 This inequality (\ref{ineq1}) shows that the presence and 
 absence of boundary interactions affect the free energy only 
 by a term proportional to $N_b$ and thus
 can be neglected in the thermodynamic limit where the leading term
 is of order $N$.

 Let us first notice that the derivative of $h(x)$ is non-negative,
 the first Griffiths inequality,
\be
 h'(x)=\lambda \frac{\Tr \sum_{\langle ij \rangle}^{(2)}S_i S_j 
 e^{-H(x)}}{\Tr e^{-H(x)}}\ge 0,
 \label{derivative1}
\ee
 where $-H(x)$ is the effective Hamiltonian appearing 
 in the exponent of (\ref{free-energy-h}).
 The denominator of (\ref{derivative1}) is positive.
 The numerator is also non-negative for $0\le x \le 1$:
 Each term of the expansion of the numerator
\be
 \Tr \sum_{\langle ij\rangle}^{(2)}S_i S_j 
 \sum_{n=0}^{\infty}\frac{\lambda^n}{n!}
 \left( \sum_{\langle ij\rangle}^{(1)}S_i S_j
 +x\sum_{\langle ij\rangle}^{(2)}S_i S_j\right)^n
 e^{-z\sum_i S_i^2}
\ee
 is composed of integrals of the form
\be
 \Tr S_i^a S_j^b S_k^c \, e^{-z\sum S_i^2},
\ee
 which is zero (if any one of $a, b, c, \cdots$ is odd) 
 or positive (otherwise).
 The second derivative is also non-negative:
\be
 h''(x)=\lambda^2 \left[\left\langle 
 \left(\sum_{\langle ij\rangle}^{(2)}S_i S_j\right)^2\right\rangle_{\rm G}
 -\left(\left\langle \sum_{\langle ij\rangle}^{(2)}S_i S_j 
 \right\rangle_{\rm G}\right)^2 \right]\ge 0,
\ee
 where $\langle \cdots \rangle_{\rm G}$ stands for 
 the average by the weight $e^{-H(x)}$.
 Thus $h'(x)$ is non-decreasing and is bounded 
 by $h'(1) (\ge 0)$ for $0\le x \le 1$.
 Therefore
\be
  |h(1)-h(0)|=\left| \int_0^1 dx \frac{dh}{dx}\right|
  \le \int_0^1 dx \left|\frac{dh}{dx}\right|
  \le h'(1). \label{ineq2}
\ee
 Our task is to upper-bound $h'(1)$.
 From the definition of $h(x)$, this derivative is expressed as
\be
 h'(1)=\lambda N_b \langle S_i S_j \rangle_{\rm G} (x=1),
\ee
 where $\langle ij \rangle$ is a bond across the interface.
 According to (\ref{ineq2}),
 if we are able to prove that $\langle S_i S_j \rangle_{\rm G} (x=1)$ is finite
 in the thermodynamic limit ($N\to\infty, N_b\to\infty$),
 we will have finished the proof that $h(1)$ and $h(0)$ are no more different
 than a quantity of order $N_b$.
 This implies that the contribution of the interface interactions
 can be neglected in the computation of the bulk free energy.

 Finiteness of $r(y=0)\equiv \langle S_i S_j\rangle_{\rm G} (x=1)$ 
 can be shown as follows.
 The argument $y$ of $r(y)$ stands for the strength of interactions
 connecting the left-most sites and right-most sites 
 along the $x$ direction.
 In other words, $y=1$ corresponds to the periodic boundary and 
 $y=0$ is for free boundary in the $x$ direction 
 (Remember that $x=1$ ensures that 
 the interactions across the interface exist).
 All other directions have periodic boundary conditions.
 The Hamiltonian is modified as
\be
 -H(x=1, y)=-z\sum_i S_i^2j+\lambda \sum_{\langle ij\rangle}^{(1)}S_i S_j
 +\lambda \sum_{\langle ij\rangle}^{(2)}S_i S_j
 +y\lambda \sum_{\langle ij\rangle}^{(3)}S_i S_j,
\ee
 where the final sum with superscript (3) runs over the boundary bonds.
 Let us assume for the moment that we have proved the following inequality,
\be
 0\le r(y=0)\le r(y=1). \label{r_ineq}
\ee
 Since $r(1)$ is the single-bond correlation 
 for fully-periodic boundary conditions,
 we can calculate it explicitly by taking the derivative of the free energy
 with respect to $\lambda$ and diving the result by the total number of bonds.
 The explicit form is available for this quantity in (\ref{Z}) and it is
 easy to see that $r(1)$ is positive and finite provided that $z>d$.
 This ends the proof that $r(0)$ is finite.

 To prove (\ref{r_ineq}), we first notice $r(0)\ge 0$, the first Griffiths
 inequality, which can be proved as we did above.
 Next we take the derivative of $r(y)$,
\be
 r'(y)=\lambda \sum_{\langle lm\rangle}^{(3)} 
 \left(\langle S_iS_jS_l S_m\rangle_{\rm G}
 -\langle S_iS_j\rangle_{\rm G} \langle S_l S_m\rangle_{\rm G}
 \right). \label{r_derivative}
\ee
 The definition of $\langle \cdots \rangle_{\rm G}$ is slightly modified 
 in that the Hamiltonian $-H(x=1, y)$ is now used.
 Since the integral defining $\langle \cdots \rangle_{\rm G}$ is Gaussian,
 Wick's theorem applies,
\be
 & & \langle S_iS_jS_l S_m\rangle_{\rm G}-\langle S_iS_j\rangle_{\rm G} 
 \langle S_l S_m\rangle_{\rm G}
 \no\\
 & & 
 = \langle S_iS_l\rangle_{\rm G} \langle S_j S_m\rangle_{\rm G}
 +\langle S_iS_m\rangle_{\rm G} \langle S_j S_l\rangle_{\rm G} \ge 0,
\ee
 due to the first Griffiths inequality.
 The proof of (\ref{r_ineq}) thus completes.

 Finally, we show that the result applies also to the case where the two
 subsystems have different values of $\lambda$.
 Let us replace $\lambda$ by $u\lambda~(0<u\le 1)$ 
 for one of the two subsystems.  
 The other subsystem keeps the original value of $\lambda$.
 Then, $r$ is a function of $y$ and $u$.
 The derivative of $r(y,u)$ with respect to $u$ has an expression
 very similar to (\ref{r_derivative}), 
 which can be shown to be positive as before.
 Thus, $r(y,u)\le r(y,1)$ for $0<u\le 1$.
 Since $u=1$ is for the system already treated above,
 we know that $r(1,1)$ is finite.
 It then follows that $r(0,u)(\le r(0,1)\le r(1,1))$ is finite.
 All other parts of the proof can trivially be generalized to
 accommodate $0<u\le 1$. Q.E.D.

 The condition of the outer boundary (free or periodic) along the $x$ axis
 can also be shown to be irrelevant in the thermodynamic limit.
 To outline the process, let us define
\be
 j(y)=\ln \Tr e^{-H(x=1,y)}.
\ee
 The goal is to prove
\be
 |j(1)-j(0)|\le c N_b,
\ee
 where $c$ is a quantity that converges to a finite value 
 in the thermodynamic limit and $N_b$ is the number of bonds 
 appearing in the  summation with superscript (3).
 To show this, according to our experience above, 
 we should prove the relations
\be
 j'(1)=cN_b,\quad j'(y)\ge 0.
\ee
 These can be proved in the same manner as before.


\end{document}